\documentclass[10pt,letterpaper]{article}

\pdfoutput=1

\usepackage[pdftex]{graphicx}
\usepackage[latin1]{inputenc}
\usepackage{amsfonts}
\usepackage{amsmath}
\usepackage{opex3}
\usepackage{cite}

\begin{document}

\title{An integrated source of broadband quadrature squeezed light}

\author{Ulrich B. Hoff,$^{1,*}$ Bo M. Nielsen,$^1$ and  Ulrik L. Andersen$^1$}

\address{$^1$ Department of Physics, Technical University of Denmark, Fysikvej bld. 309, 2800 Kgs. Lyngby, Denmark}

\email{$^*$ulrich.hoff@fysik.dtu.dk}

\begin{abstract}
An integrated silicon nitride resonator is proposed as an ultra-compact source of bright single-mode quadrature squeezed light at 850\,nm. Optical properties of the device are investigated and tailored through numerical simulations, with particular attention paid to loss associated with interfacing the device. An asymmetric double layer stack waveguide geometry with inverse vertical tapers is proposed for efficient and robust fibre-chip coupling, yielding a simulated total loss of -0.75\,dB/facet. We assess the feasibility of the device through a full quantum noise analysis and derive the output squeezing spectrum for intra-cavity pump \emph{self-phase modulation}. Subject to standard material loss and detection efficiencies, we find that the device holds promises for generating substantial quantum noise squeezing over a bandwidth exceeding 1\,GHz. In the low-propagation loss regime, approximately -7 dB squeezing is predicted for a pump power of only 50 mW.
  \end{abstract}


%%%%% BIBLIOGRAPHY %%%%%%

%%% END BIBLIOGRAPHY %%%%%%%%%%

\section{Introduction}

Over the last decades an immense progress in control and manipulation of quantum states of light and matter and their mutual interaction has facilitated generation of squeezed states of light in a broad selection of physical systems: from the pioneering works on four-wave mixing in atomic vapors~\cite{Slusher1985} and optical fibres~\cite{Shelby1986} over commonly used workhorse systems based on spontaneous parametric down-conversion in $\chi^{(2)}$-materials to recent demonstrations of ponderomotive squeezing through optomechanical interactions~\cite{Brooks2012,Purdy2013,Safavi-Naeini2013}.

In the emergent field of optical quantum technologies, squeezed quantum states of light have become an essential resource for many continuous variable (cv) quantum information and communication protocols~\cite{Madsen2012}. And the feasibility of exploiting such quantum-correlated states for ultra-sensitive measurements has been demonstrated in numerous implementations of quantum-enhanced sensing, pioneered in the context of Mech-Zehnder interferometry~\cite{Caves1981,Xiao1987}, and by now applied at both micro- and macroscopic scales~\cite{Hoff2013,LIGO2011}. However, the size and operational complexity of quantum light sources remains a limitation for practical and industrial applications of optical quantum technologies and integration with electronics. Efficient pulsed Kerr-squeezing in standard polarization maintaining fibres~\cite{Bergman1991,Heersink2005} requires fibre lengths on the order of tens of meters and even monolithic squeezed light sources~\cite{Kurz1993,Breitenbach1995,Eberle2010} employ bulk nonlinear crystals that are orders of magnitude larger than lithographically defined electronic integrated circuits. Generation of twin beam squeezing has been demonstrated in whispering gallery mode disk resonators~\cite{Furst2011} and recently in resonant on-chip structures~\cite{Dutt2013}, bringing the development of a fully integrated opto-electronic quantum sensing device an important step closer. But so far, an integrated source of quantum correlated cv states remains to be demonstrated.  

One major limitation for the integration of quantum light sources is optical loss. Free-space setups mainly suffer from Fresnel losses at optical interfaces, which can be mitigated by tailored anti-reflection coatings, reducing the effect to the 0.1\% level per interface. However, in integrated waveguide circuits the light is also subject to material losses and scattering losses, originating from the roughness of the lithographically defined sidewalls and increasing quadratically with the refractive index contrast $\Delta n = n_{core}-n_{cladding}$~\cite{Daldosso2004}. The former contribution is governed by intrinsic properties of the material's band structure and cannot be compensated for. Bulk crystalline silicon has a bandgap of 1.1\,eV leading to significant two-photon absorption (TPA) at wavelengths below roughly 2000\,nm,  and this is the reason why so much attention is currently drawn towards identification of CMOS-compatible alternatives to the otherwise so appealing silicon-on-insulator (SOI) platform for nonlinear optics. Scattering losses, on the other hand, can be compensated for by optimization of fabrication procedures and engineering of waveguide dimensions.  In stoichiometric silicon nitride films deposited with low-pressure chemical vapor deposition (LPCVD),  200\,nm thick channel waveguides with $\leq \!-0.2$ dB/cm propagation loss at 780\,nm have been demonstrated~\cite{Daldosso2004}. Using temperature-cycling to overcome the tensile-stress limited film thickness of around 250\,nm, propagation losses of -0.12 dB/cm at 1540\,nm have been measured in channel waveguides of more than 700\,nm thickness~\cite{Gondarenko2009}. And by employing a large aspect-ratio waveguide geometry the dominating scattering loss contribution can be shifted from the sidewalls to the waveguide top and bottom surfaces, which for LPCVD ${\rm Si}_3{\rm N}_4$ have roughnesses in the sub-nanometer regime, resulting in ultra-low waveguide propagation losses below -0.1\,dB/m at 1580\,nm~\cite{Bauters2011}. Unfortunately, this comes at the price of low field confinement, which is essential for efficient nonlinear processes. An alternative CMOS-compatible material is high-index doped silica (Hydex) which has recently been demonstrated to be a promising candidate for integrated nonlinear optics~\cite{Moss2013}. Due to its high linear refractive index ($n$ = 1.5 - 1.9) and low propagation loss of $ \sim {\rm -0.06\,dB/cm}$ a nonlinear parameter of $\gamma \sim 233\,{\rm W^{-1}km^{-1}}$ at 1550\,nm is achievable in high-confinement channel waveguides~\cite{Ferrera2008}.

Yet another severe loss-related challenge for high-index contrast waveguides is the inevitable mode conversion and effective index step associated with chip interfacing. The former leads to loss through coupling to radiation modes and the latter to Fresnel loss and low collection efficiency because of the large numerical aperture of high-index contrast waveguides. The most common approach to overcome these complications is to employ inverse tapers (spot size converters) at the waveguide ends allowing near-adiabatic conversion between the tightly confined waveguide mode and that of a single mode tapered fibre or fibre pigtail. Using such coupling strategies approximately -1.5 dB/facet coupling loss has been reported for pigtailed high-index contrast Hydex waveguides~\cite{Ferrera2008}.   

In this paper, we study theoretically the feasibility of an integrated silicon nitride waveguide resonator as a source of quadrature squeezed light around 850 nm, conveniently centered in the Ti:sapphire tuning range. Classical optical parametric oscillations~\cite{Levy2009} and  frequency comb generation~\cite{Okawachi2011} have been demonstrated in similar structures at telecom wavelengths, requiring careful engineering of dispersion properties in order to achieve a broad phase matching bandwidth. For the application discussed here, the phasematching condition is relaxed because the quantum correlated fields of interest are co-resonating rf-sidebands symmetrically distributed around the pump. This, however, introduces further complications at the state-interrogation stage, since the sidebands cannot easily be separated from the bright pump field and the phase-space rotations required for standard homodyne state tomography are not readily realizable. Integrated silicon nitride ring resonators have also been proven viable sources of photon pairs with controlable degree of correlation~\cite{Helt2011}. But so far, the cv quantum noise properties of the parametrically generated sideband fields have remained unexplored experimentally. Optical losses are detrimental to quantum correlated states in general but quadrature squeezed states are particularly vulnerable. The effect of optical loss on such states is not just a mere reduction in production rate but an over all degradation of quantum correlations and thereby it's practical applicability. For a given amount of loss the state degradation is more pronounced for larger degrees of squeezing, rendering moderate squeezing levels a more practically tractable target.

The primary objective of our analysis is to identify a feasible design for such an integrated device, subject to the constraints of conventional microfabrication techniques and commonly achievable propagation loss figures. We address this task by performing numerical simulations of optical key properties of the waveguide structure and using the results as inputs to a full quantum noise analysis of the nonlinear intra-cavity interaction. Special attention is paid to optimization of chip in and out coupling since this is in general a bottleneck for the integration of quantum light sources. 

\section{Silicon nitride}
Silicon nitride is a CMOS compatible material widely used in microelectronics, both as gate dielectric in thin-film transistors and as mask, due to its high etching selectivity over silicon. In the field of integrated optics LPCVD deposited amorphous stoichiometric silicon nitride (${\rm Si}_3{\rm N}_4$) has only recently been introduced as a platform for nonlinear optics but has already been intensively investigated. Due to its linear refractive index of $n\approx 2$ it is a suitable core material for high-index contrast waveguides in applications where tight field confinement and small waveguide bending radii are required. With a bandgap of approximately 5\,eV~\cite{Ghodssi2011}, the absorption edge of silicon nitride is just below 300\,nm, providing low material loss in the visible and infrared, and above 600\,nm TPA does not represent a limitation even at high power levels, as opposed to silicon.  For the particular application at hand, an equally important property of silicon nitride is the reasonably high third order Kerr nonlinearity $n_2 = 2.5 \times 10^{-15} \rm{cm^2/W}$~\cite{Ikeda2008}, which is essential for implementation of parametric processes, enabling manipulation of the quantum properties of light. For comparison, the nonlinearity of silicon nitride is roughly 10 times the nonlinearity of silica, 2 times that of Hydex, and 0.1 times that of silicon. Thus, with sufficient optical isolation from the silicon substrate, e.g. using silicon dioxide as cladding material, silicon nitride is a promising candidate for integration of quantum optical circuits where minimization of losses and strong interactions are of uttermost importance.

\section{The system}
\label{sec: the system}
\begin{figure}[htbp!]
\centering
\includegraphics[width=0.83\columnwidth]{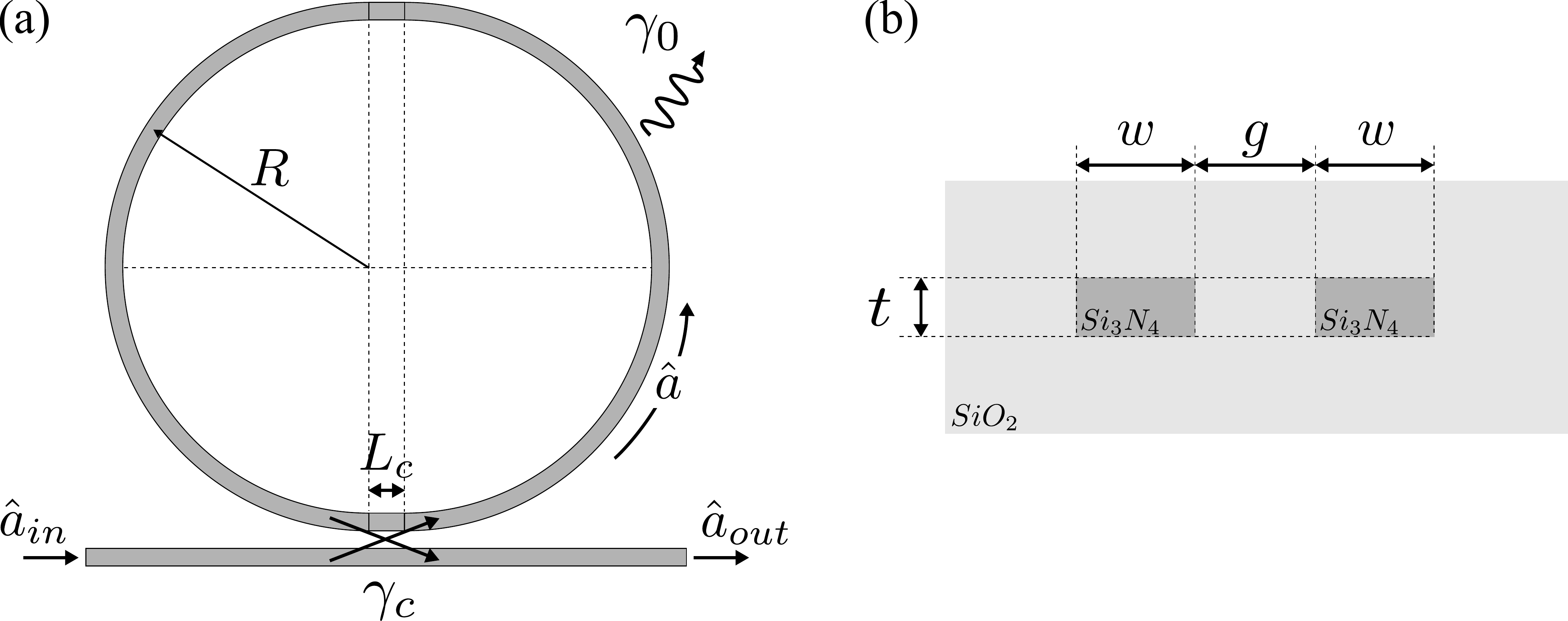}
\caption{(a) Top view of the racetrack resonator geometry. (b) Cross sectional view at the bus-resonator coupling region. The thickness is $t  = 250\,{\rm nm}$ and the width $w$ is chosen such that the waveguide only supports a single transversal mode. The gap size $g$ is fixed, and limited from below by the resolution of the lithographic method chosen, and the coupling rate $\gamma_{\rm c}$ is determined by the length of the coupling region $L_{\rm c}$.}
\label{Fig: RTR_topview}
\end{figure}

The specific integrated system that we are concerned with in this paper is a buried channel waveguide circuit consisting of a racetrack resonator (RTR) with radius of curvature $R$, laterally coupled to a straight bus waveguide  (Fig.~\ref{Fig: RTR_topview}). A finite overlap of the evanescent fields of the two waveguide modes in the coupling region allows coupling to the RTR with an efficiency modelled by the field amplitude coupling rate $\gamma_{\rm c} \approx \kappa_{\rm c}^2/2\tau$
%\footnote{The integrated resonator can be modelled by an equivalent one-sided bulk-optics cavity with internal loss, for which  $\kappa_c^2$ is the intensity transmittivity of the coupling mirror. The approximate expression is valid for $\kappa_c^2 \ll 1$.}
, where $\tau = 2n_{\rm eff}(L_{\rm c} + \pi R)/c$ is the resonator round-trip time and $\kappa_c^2$ is the intensity transmittivity of the coupling mirror of an equivalent one-sided bulk-optics cavity. The approximate expression for $\gamma_{\rm c}$ is valid for $\kappa_{\rm c}^2 \ll 1$. The coupling rate is controlled through the gap size and coupling length parameters $g$ and $L_{\rm c}$, respectively. Intra-cavity losses, primarily due to scattering from the waveguide sidewalls, are represented by an intrinsic loss rate $\gamma_{\rm 0} \approx \kappa_{\rm 0}^2/2\tau = \alpha c / 2 n_{\rm eff}$, where $\alpha$ is the waveguide per meter loss parameter, $c$ is the vacuum speed of light, and $n_{\rm eff}$ the effective mode index. The total loss rate is given by  $\gamma = \gamma_{\rm 0} + \gamma_{\rm c}$ (HWHM) determining the loaded quality factor of the resonator. An important parameter for controlling the squeezed light generation efficiency of the system is the escape efficiency $\eta_{\rm esc} = \gamma_{\rm c}/\gamma$, characterizing the collection efficiency of the intra-cavity field. The escape efficiency determines the coupling regime with $\eta_{\rm esc} = 1/2$ corresponding to critical coupling and $\eta_{\rm esc} < 1/2$ ($\eta_{\rm esc} > 1/2$) to under (over) coupling. For the application of squeezed light generation, operation in the over coupled regime is required, but obviously there is a trade-off between efficient collection from the cavity (large $\eta_{\rm esc}$) and resonant enhancement of the nonlinear process (small $\gamma$). 

As depicted in Fig.~\ref{Fig: RTR_topview}(b), we opt for a rectangular low-aspect ratio cross sectional geometry of the waveguides. For the sake of minimizing fabricational complexity, we restrict the thicknesss $t$ to a value of 250\,nm, yielding the largest possible mode confinement while at the same time eliminating the need for stress-releasing temperature cycling in the deposition process. This in turn puts an upper bound on the width $w$, due to the requirement of single mode operation (cf. section~\ref{Sec: waveguide mode properties}).

\section{Kerr nonlinearity}
\label{sec: Kerr nonlinearity}
The underlying mechanism for nonlinear optical effects in Kerr media is the third order tensor susceptibility $\chi^{(3)}_{ijkl}$, supporting a great variety of four-photon mixing processes, collectively termed \emph{four-wave mixing} (FWM). In general, FWM processes involve conversion of two pump photons ($\omega_{\rm p}$) into a pair of photons at signal and idler frequencies ($\omega_s$,\, $\omega_i$), respectively, with the convention that $\omega_{\rm s} > \omega_{\rm i}$. The interacting field frequencies are related by $2\omega_{\rm p} = \omega_{\rm s} + \omega_{\rm i}$, required by energy conservation, and in order for the process to occur efficienctly the phase matching condition $\Delta \beta =2 \beta_{\rm p} - \beta_{\rm s} - \beta_{\rm  i} \approx 0$ must be fulfilled; here expressed in terms of the guided mode propagation constants  $\beta_{i} = n_{\rm eff}(\omega_{i}) k_{0,i}$ where $n_{\rm eff}$ is the effective mode index. For this reason, the dominant interaction is the intrinsically phase matched \emph{self-phase modulation} (SPM) process corresponding to the case of complete degeneracy of the partaking fields and resulting in an intensity dependent refractive index $n(I) = n_0 + n_2 I$. In this case the pump field experiences a self-induced phase shift  $\phi_{\rm nl}= 2\pi n_2 I L/\lambda = \gamma_{\rm nl} L P_{\rm p}$, where $\gamma_{\rm nl}=\omega n_2 /c A_{\rm eff}$ is the nonlinear parameter, $A_{\rm eff}$ the effective mode area, and $P_{\rm p}$ the pump power. It is well-known that this type of FWM is capable of generating self-induced single-mode squeezing~\cite{Bachor2004,Sizmann1999} and it is also the interaction of primary interest to this work. However, in the context of nonlinear optics in integrated resonators a more commonly discussed interaction is \emph{non-degenerate four-wave mixing} (NDFWM) where an intense pump field interacts with two or a multiple of weaker non-degenerate signal and idler fields, each of the fields being resonant on separate longitudinal resonator modes. Parametric oscillators~\cite{Levy2009,Razzari2009} and Kerr-frequency combs~\cite{Herr2012} have previously been demonstrated using this interaction, and generation of squeezing, both below and above threshold, is discussed in~\cite{Jack1995}.

A simple and intuitive understanding of pump squeezing by SPM is provided in~\cite{Bachor2004}: Consider an input pump field $\alpha = \alpha_0 + \delta \!X_1^{\rm in} + \delta\!X_2^{\rm in}$ consisting of a coherent carrier amplitude and vacuum fluctuations in the sideband quadratures defined as $\delta\!X_{\theta} =  \delta a e^{-i\theta} + \delta a^{\dagger} e^{i\theta}$ with $\delta\!X_1 = \delta\!X_{\theta=0}$ and $\delta\!X_2 = \delta\!X_{\theta=\pi/2}$. Both the coherent amplitude $\alpha_0$ and the noise $\delta\!X_1^{in}$ drives the SPM process, resulting in a phase shift of the carrier proportional to $\alpha_0$ and induced phase fluctuations at sideband frequencies given by $\delta \phi_{\rm nl} = 4\pi n_2 L \alpha_0 \delta\!X_1^{\rm in}/\lambda$. For small angles the corresponding change in the phase quadrature fluctuations is $\Delta(\delta\!X_2) \approx \alpha_0 \delta\phi_{\rm nl} \equiv 2 r \delta\!X_1^{\rm in}$, yielding the following input-output relations for the quadrature fluctuations under influence of SPM:
\begin{eqnarray}
\delta\!X_1^{\rm out} &=& \delta\!X_1^{\rm in} \\
\delta\!X_2^{\rm out} &=& \delta\!X_2^{\rm in} + 2 r \delta\!X_1^{\rm in}
\label{eq: Kerr input-output}
\end{eqnarray}
While the amplitude quadrature is unaffected by the interaction, an amount of the amplitude fluctuations proportional to the carrier intensity is admixed to the phase quadrature thereby correlating the two. The correlations are strongest at a quadrature phase of $\theta_{\rm min}(r) = \frac{1}{2}\arctan (-1/r)$ for which the variance is ${\rm Var}(\delta\!X_{\theta_{\rm min}}^{\rm out}) = 1 - 2r\sqrt{1+r^2} + 2r^2$. Thus, for any $r>0$ the noise is, in principle, squeezed below the vacuum level at the optimal phase angle which in the limit of weak interaction strength is $\lim_{r \rightarrow 0} \theta_{\rm min}(r) = -\pi/4$. However, the efficiency of the third order nonlinearity is very low and appreciable $r$-values are only achievable for very strong pump fields. In case of cw pumping this requirement can be addressed by resonant enhancement of the interaction which will be our focus in the following section.

\section{Quantum dynamics of resonantly enhanced SPM}
\label{sec: quantum dynamics}
In this section we study the effects of SPM on the quantum dynamics and noise properties of a cavity field interacting with a Kerr nonlinearity. Quantum mechanically the interaction is governed by the quartic term 
\begin{equation}
H_{\rm SPM} = \frac{\hbar \xi}{2}{a^{\dagger}_{\rm p}}^2 a_{\rm p}^2,
\label{eq: spm Hamiltonian}
\end{equation}
of the full FWM Hamiltonian (see Appendix \ref{ap: Kerr Hamiltonian}), where the interaction strength is given by
\begin{equation}
\xi = \frac{\hbar \omega c^2 \gamma_{\rm nl}}{2 n_{\rm eff}^2 L}.
\label{eq: coupling constant}
\end{equation}
Following the quantum Langevin approach we formulate linearized equations of motion for the intracavity quantum fluctuations, and the single mode squeezing spectrum for the output field is derived. 

\subsection{Langevin equation of motion}
Considering a single pumped resonator mode $a_{\rm p}$ interacting with a Kerr nonlinearity the general Langevin equation of motion is given by \cite{Walls1995,Gardiner2000}
\begin{equation}
\frac{d a_{\rm p}}{d t} = -\frac{i}{\hbar} \left[ a_{\rm p}, H \right] -\gamma a_{\rm p} + \sqrt{2 \gamma_{\rm c}} a_{\rm p,in}e^{-i\omega_L t} + \sqrt{2 \gamma_{\rm 0}} b_{\rm p},
\end{equation}
where  
\begin{equation}
H = \hbar \omega_{\rm p} {a_{\rm p}}^{\dagger}a_{\rm p} + H_{\rm SPM}.
\end{equation}
Operators $a_{\rm p,in}$ and $b_{\rm p}$ represent driving and input vacuum fields, respectively. Transforming all mode operators to a frame rotating at the driving laser frequency $\omega_{\rm L}$, by means of the substitution $a_{\rm p} \rightarrow e^{-i\omega_L t}A_{\rm p}$, we get the rotating frame equation of motion:
\begin{equation}
\frac{d A_{\rm p}}{d t} = 
-(\gamma - i\Delta_{\rm p})A_{\rm p} 
-i\xi A_{\rm p}^{\dagger} A_{\rm p}^2
+\sqrt{2 \gamma_{\rm c}} A_{\rm p,in} 
+\sqrt{2 \gamma_0} B_{\rm p},
\label{eqn: Langevin pump}
\end{equation}
with detuning $\Delta_{\rm p} = \omega_{\rm L} - \omega_{\rm p}$ relative to the empty cavity mode frequency. Adapting the standard coupled-mode theory normalization convention, $|A_{\rm p}|^2$ represents the total mode photon number stored in the resonator and $|A_{\rm p,in}|^2$ represents the traveling mode photon flux.  

\subsection{Linearized dynamics}
In order to solve the system dynamics we linearize the rotating frame quantum Langevin equation Eq.~(\ref{eqn: Langevin pump}), about the solution $\alpha_p$ to the steady-state equation 
\begin{equation}
|\alpha_{\rm p}|^2 (\gamma^2 + (\Delta_{\rm p} - \xi |\alpha_{\rm p}|^2)^2) = 2\gamma_{\rm c} |\alpha_{\rm p,in}|^2.
\label{eqn: bistability}
\end{equation}
We do so by making the substitution $A_{\rm p} \rightarrow \alpha_{\rm p} + \delta a_{\rm p}$ and retaining only terms of first order in the quantum fluctuation operator $\delta a_{\rm p}$.
Defining the vector fluctuation operators $\mathbf{\delta a}_{\rm i} = (\delta a_{\rm i},\, \delta a_{\rm i}^{\dagger})^{\top}$ the linearized system of differential equations can be stated as a matrix equation:
\begin{equation}
\frac{d \mathbf{\delta a}_{\rm p}}{d t} = \left[ \mathbf{M} - \gamma \mathbf{I} \right] \mathbf{\delta a}_{\rm p} + \sqrt{2\gamma_{\rm c}} \mathbf{\delta a_{\rm in}} + \sqrt{2\gamma_0} \mathbf{\delta b},
\label{eqn: linearized qle}
\end{equation}
where the dynamics of the pumped mode is fully characterized by the system matrix
\begin{equation}
\mathbf{M}-\gamma \mathbf{I}=\left(
\begin{array}{cc}
-\gamma -i(2|\epsilon|-\Delta_{\rm p}) & -i\epsilon \\
i\epsilon^* & -\gamma + i(2|\epsilon| - \Delta_{\rm p})
\label{eqn: spm system matrix}
\end{array}
\right),
\end{equation}
and its eigenvalues
\begin{equation}
\lambda_{\pm}=-\gamma \pm \sqrt{|\epsilon|^2-(\Delta_{\rm p} - 2|\epsilon|)^2}.
\label{eqn: pump eigenvalues}
\end{equation}
For brevity we have introduced the pump parameter $\epsilon = \xi |\alpha_{\rm p}|^2 e^{i2\phi}$, where $\phi$ is the phase of the pump field. From Eq.~(\ref{eqn: spm system matrix}) we observe that SPM induces phase-dependent correlations in the fluctuations of the intra-cavity field (anti-diagonal elements), increasing with pump power, and a power-dependent nonlinear shift of the cold-cavity resonances (diagonal elements), responsible for optical bistability of the system. 

\begin{figure}[htbp!]
\centering
\includegraphics[width=0.75\columnwidth]{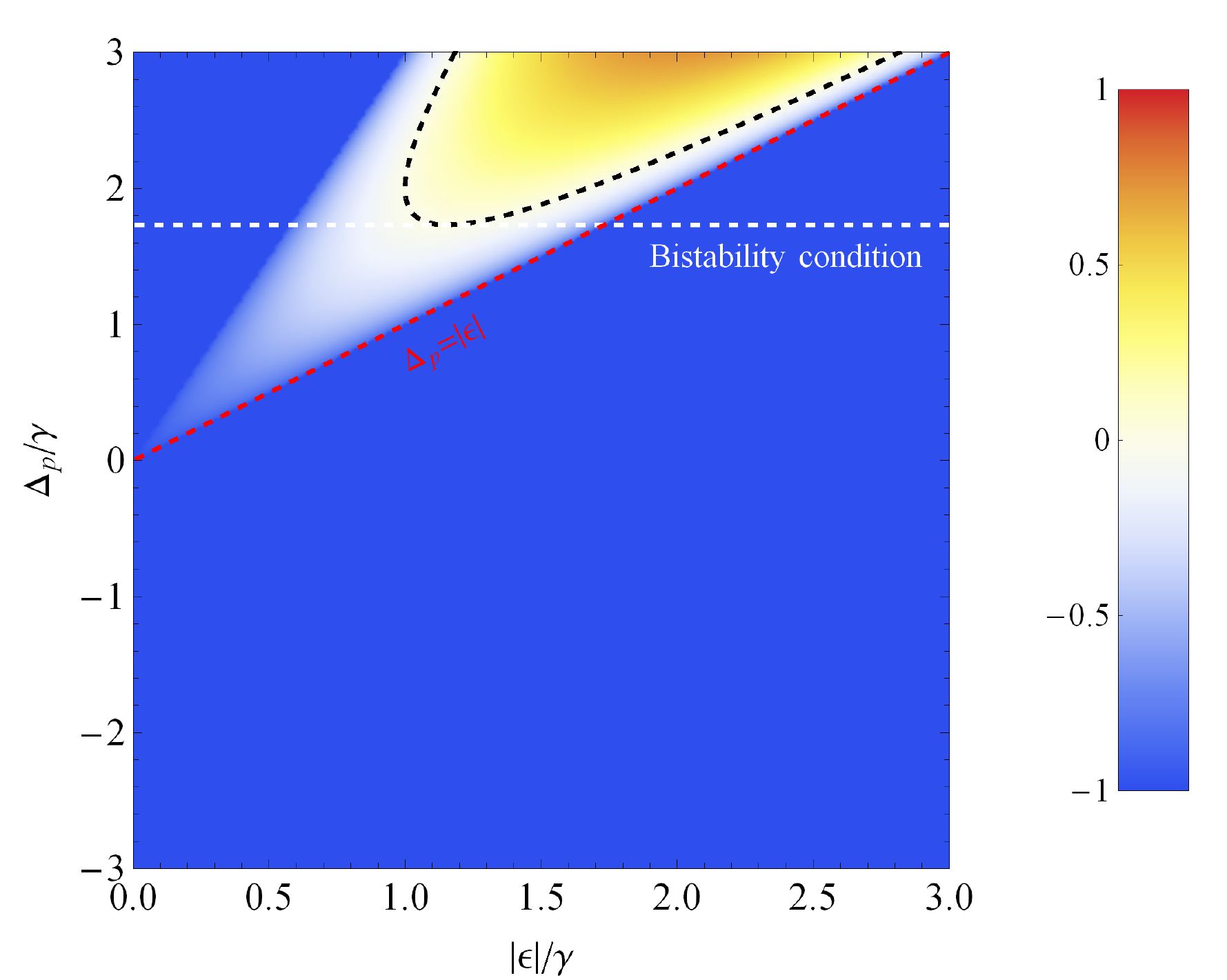}
\caption{Real part of the pumped mode eigenvalue $\lambda_+$ as a function of pump parameter and detuning. The $\lambda_+ = 0$ contour (black dashed line) divides the parameter space into a stable ($\lambda_+ <0$) and an unstable  ($\lambda_+ >0$) region. For detunings above the bistability condition (white dashed line) two stable regions exist and the system can be tuned between the two by changing the intracavity power. The intermediate unstable region is not accessible. By progressively detuning the drive field from the empty cavity resonance as the pump power is increased it is possible to maintain stability of the system and maximize the intra-cavity power (red dashed line). Furthermore, we observe that in terms of power the instability region is bounded from below by the condition that $|\varepsilon|>\gamma$ indicating that the nonlinear scattering rate from the pumped mode should exceed the cavity dissipation rate in order for instability to occur.}
\label{Fig: PumpEigenvalue}
\end{figure}

Stability of the pumped mode requires that ${\rm Re}(\lambda)<0$ for all eigenvalues. This is trivially fulfilled for $\lambda_-$, and for $\lambda_+$ the condition can be equivalently formulated as: $3 |\epsilon|^2 -4 \Delta_{\rm p} |\epsilon| +\Delta_{\rm p}^2 +\gamma^2 > 0$. The real part of $\lambda_+$ is plotted in Fig.~\ref{Fig: PumpEigenvalue}.
Solving the equation for $|\epsilon|$ we find solutions $|\epsilon|_{\pm}=\frac{1}{3}(2\Delta_{\rm p} \pm \sqrt{\Delta_{\rm p}^2 -3\gamma^2})$; for $|\epsilon|_- \leq |\epsilon| \leq |\epsilon|_+$  the system is unstable and outside it is always stable. Furthermore, for $\Delta_{\rm p} <\sqrt{3} \gamma$ no real positive solution for $|\epsilon|$ exist and the stability criterion is always fulfilled; for $\Delta_{\rm p} >\sqrt{3} \gamma$ two solutions exist, marking the onset of optical bistability. We observe from Eq.~(\ref{eqn: pump eigenvalues}) that for the particular choice of detuning $\Delta_{\rm p} = \xi|\alpha_{\rm p}|^2 = |\epsilon|$ the eigenvalues collapse to a degenerate pair  $\lambda_{\pm}=-\gamma$. Thus, the system is always stable, independent of the pump power. In this case the SPM induced nonlinear shift of the cavity resonances is exactly compensated for by the pump laser detuning, restoring the usual proportionality between input pump power and stored intra-cavity power, $|\alpha_{\rm p}|^2 = 2\gamma_{\rm c}/\gamma^2 \cdot |\alpha_{\rm p, in}|^2$ for a resonantly driven cavity.

\subsection{Squeezing spectrum}
Since we are ultimately interested in deriving the squeezing spectrum for the system it is convenient to transform Eq.~(\ref{eqn: linearized qle}) into frequency space. Fourier transformation of the differential equation system yields the following set of algebraic equations,
\begin{equation}
-i\Omega \, \mathbf{\delta a}_{\rm p}(\Omega) = \left[ \mathbf{M} - \gamma \mathbf{I} \right] \mathbf{\delta a}_{\rm p}(\Omega) + \sqrt{2\gamma_{\rm c}}\, \mathbf{\delta a_{\rm in}}(\Omega) + \sqrt{2\gamma_{\rm 0}}\, \mathbf{\delta b}(\Omega),
\end{equation}
which can readily be solved for the intra-cavity fluctuations:
\begin{equation}
\mathbf{\delta a}_{\rm p}(\Omega) = -\left[ \mathbf{M} -(\gamma -i\Omega)\mathbf{I} \right]^{-1} \left( \sqrt{2\gamma_{\rm c}} \mathbf{\delta a_{\rm in}} (\Omega) + \sqrt{2\gamma_0} \mathbf{\delta b}(\Omega) \right).
\label{eqn: intracavity fluc}
\end{equation}
The intra-cavity field fluctuations can in turn be projected onto a set of output modes of the cavity using the input-output formalism. In this particular case the appropriate boundary condition is,
\begin{equation}
\delta \mathbf{a}_{\rm out}(\Omega) = \delta \mathbf{a}_{\rm in}(\Omega) - \sqrt{2\gamma_{\rm c}} \delta \mathbf{a}_{\rm p}(\Omega).
\label{eqn: input-output}
\end{equation}
Inserting Eq.~(\ref{eqn: intracavity fluc}) into Eq.~(\ref{eqn: input-output}) we find
\begin{eqnarray}
\mathbf{\delta a_{\rm out}}(\Omega) &=& \left[ \mathbf{M} -(\gamma - i\Omega)\mathbf{I} \right]^{-1} \nonumber \\
&& \qquad \times \Big( \left[ \mathbf{M} + (\Delta\gamma + i\Omega)\mathbf{I} \right] \mathbf{\delta a_{\rm in}}(\Omega) + 2\sqrt{\gamma_0 \gamma_{\rm c}} \mathbf{\delta b}(\Omega)\Big),
\label{eqn: output operators}
\end{eqnarray} 
where $\Delta \gamma = \gamma_{\rm c} - \gamma_0$ characterizes the coupling regime: $\Delta \gamma < 0$ -- under coupling, $\Delta \gamma = 0$ -- critical coupling, and $\Delta \gamma > 0$ -- over coupling.

In general, an experimental interrogation of the cavity output field involves homodyne detection of the field quadratures
\begin{equation}
\delta\!X_{\theta}(\Omega) =  \delta a_{\rm out}(\Omega) e^{-i\theta} + \delta a_{\rm out}^{\dagger}(-\Omega) e^{i\theta},
\label{eqn: quadrature def}
\end{equation}
for which the quantum noise properties are fully characterized by the normal ordered second order moment 
\begin{eqnarray}
\langle :\! \delta\!X_{\theta}(\Omega) \delta\!X_{\theta}(\Omega')\!: \rangle
&=& \Big[ e^{-2i\theta}\langle\delta a_{\rm out}(\Omega) \delta a_{\rm out}(\Omega') \rangle + e^{2i\theta}\langle \delta a_{\rm out}^{\dagger}(-\Omega) \delta a_{\rm out}^{\dagger}(-\Omega') \rangle \nonumber \\
&& + \langle \delta a_{\rm out}^{\dagger}(-\Omega) \delta a_{\rm out}(\Omega') \rangle \Big] \delta (\Omega + \Omega').
\label{eqn: normal ordered moment}
\end{eqnarray}
Following~\cite{Walls1995} the normal ordered spectrum of $\delta\!X_{\theta}(\Omega)$ is then found by integrating Eq.~(\ref{eqn: normal ordered moment}) over $\Omega'$, 
\begin{equation}
:\!S_{\theta}(\Omega)\!: =\int d\Omega' \langle :\! \delta\!X_{\theta}(\Omega) \delta\!X_{\theta}(\Omega')\!: \rangle,
\label{eqn: normal ordered spectrum}
\end{equation}
and finally the squeezing spectrum of the output field quadrature is given by
\begin{equation}
S_{\theta}(\Omega) = 1 \, + :\!S_{\theta}(\Omega)\!:
\label{eqn: squeezing spectrum}
\end{equation}
which can be directly measured as the power spectral density of the homodyne photo current.  The additional constant term accounts for the vacuum variance contribution from anti-normal ordered terms of the second order moment, which for the particular field quadrature definition in Eq.~(\ref{eqn: quadrature def}) is equal to one. 

Using the SPM system matrix in Eq.~(\ref{eqn: spm system matrix}) we can evaluate the contributing second order moments in Eq.~(\ref{eqn: normal ordered moment}). Combining terms results in the following expression for the squeezing spectrum of the pumped mode:
\begin{eqnarray}
\!S^{\rm SPM}_{\theta}(\Omega)\! &=& 1\, + :\!S^{\rm SPM}_{\theta}(\Omega)\!: \nonumber \\
&=& 1 + G \cdot
\Big[ 2\gamma |\epsilon| - 2\gamma \big[ 2|\epsilon| - \Delta_{\rm p}\big]\cos 2(\theta + \phi) \nonumber \\
&& \qquad -\big[ \gamma^2 + \Omega^2 -\Delta_{\rm p}^2 + 4\Delta_{\rm p} |\epsilon| -3|\epsilon|^2 \big] \sin 2(\theta + \phi) \Big],
\label{eqn: pump spectrum}
\end{eqnarray}
where
\begin{equation}
G =  \frac{4\gamma_{\rm c} |\epsilon|}{\big[ \Delta_{\rm p}^2 + \gamma^2 -\Omega^2 - 4\Delta_{\rm p} |\epsilon| + 3|\epsilon|^2 \big]^2 + 4\gamma^2 \Omega^2}.
\end{equation}
Finite detection efficiency, primarily resulting from fibre-chip coupling ($\eta_{\rm c}$), imperfect mode overlap on the homodyne detector ($\eta_{\rm mm}$), and photodiode quantum efficiency ($\eta_{\rm qe}$), causes admixing of vacuum noise and reduces the measurable squeezing according to:
\begin{equation}
S_{\theta}^{\rm meas}(\Omega) = (1-\eta) + \eta\, S_{\theta}(\Omega) = 1 + \eta :\!S_{\theta}(\Omega)\!:,
\label{eqn: effective squeezing spectrum}
\end{equation} 
where $\eta = \eta_{\rm c} \cdot \eta_{\rm mm} \cdot \eta_{\rm qe}$ is the total detection efficiency.

\section{Numerical simulations}
\label{sec: numsim}
Up until now we have been considering the theoretical aspects of squeezed light generation in third order nonlinear materials, not accounting for actual material properties and system design. In this section we study the dependence of crucial optical propagation and nonlinear properties on geometrical system design parameters through numerical simulations. We fix the optical design wavelength to $\lambda=850\,{\rm nm}$, and the parameter space is constrained by aiming for a design with waveguide thickness $t= 250\,{\rm nm}$ and minimum lateral feature sizes $\ge\!400$\,nm. We do so in order to make the resulting design compatible with simple LPCVD single layer deposition and UV stepper lithography. All simulations have been carried out using the FieldDesigner and OptoDesigner software packages from PhoeniX Software.

\subsection{Waveguide mode properties}
\label{Sec: waveguide mode properties}
\begin{figure}[htbp!]
\centering
\includegraphics[width=0.8\columnwidth]{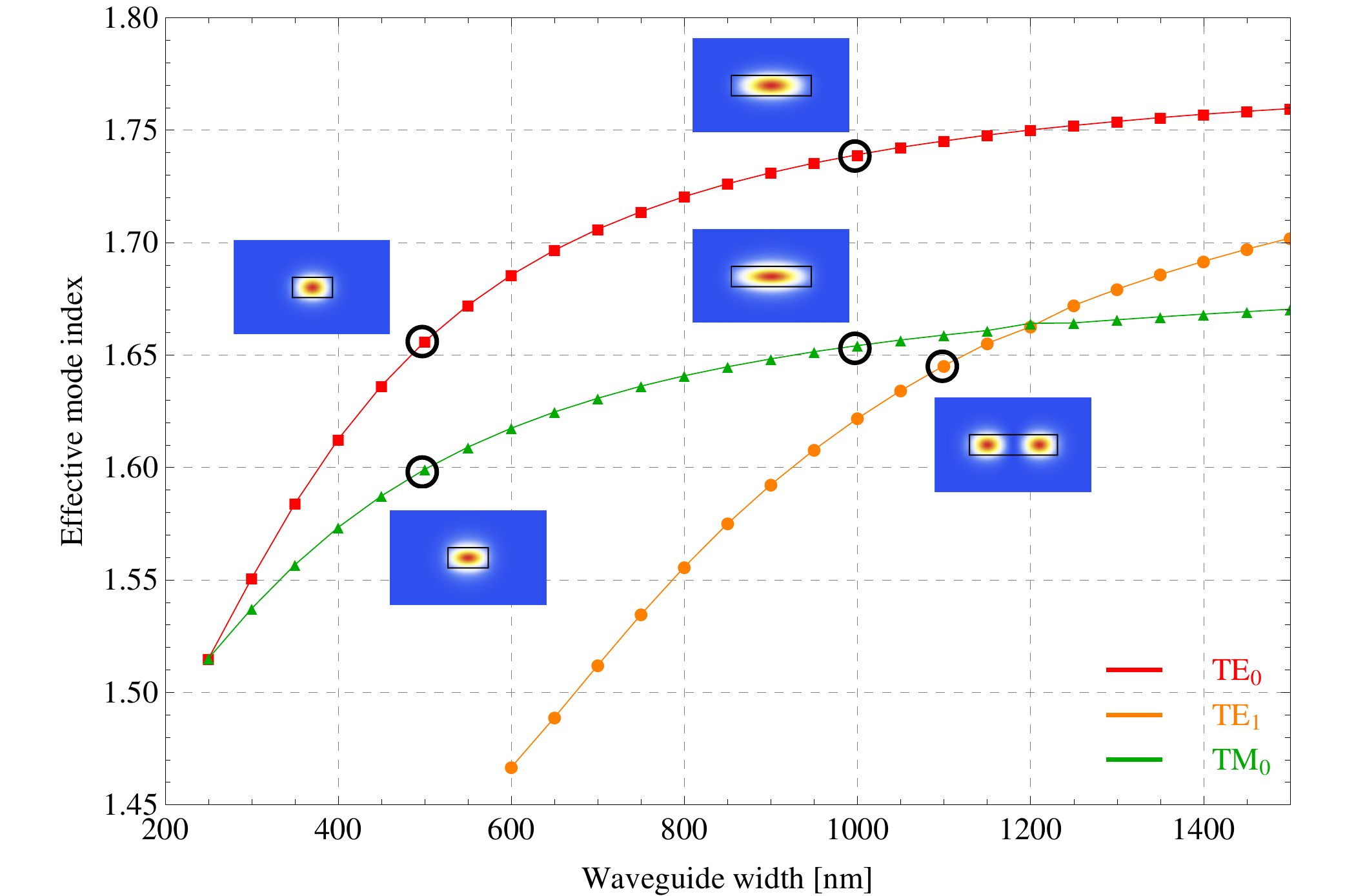}
\caption{Simulated ($\lambda=850\,{\rm nm}$, t = 250\,nm) effective mode indices for the first three guided modes as function of waveguide width. Insets: Simulated mode profiles corresponding to each of the five encircled points. Solid black lines outline the waveguide core cross section.
}
\label{Fig: Effective mode index}
\end{figure}

A fundamental requirement for the waveguide architecture is that the transversal geometry should support only a single guided mode. This is important in order to optimize power confinement in the waveguide core and to facilitate mode matching of the output field to a free space local oscillator field as required for homodyne interrogation. In Fig.~\ref{Fig: Effective mode index} we plot the effective mode indices for the first three guided modes as function of waveguide width $w$, derived from a full-vectorial mode analysis of the waveguide cross section. Starting from the square cross section and increasing the $w\!:\!t$ aspect ratio the effective indices of the fundamental $\rm TE$ and $\rm TM$ modes separate with $n_{\rm eff}^{\rm TE}>n_{\rm eff}^{\rm TM}$, indicating that the horizontally polarized TE mode is more confined to the silicon nitride core. This geometry-induced birefringence improves the polarization maintaining property of the waveguide and it is thus beneficial to maximize the aspect ratio. However, as the width is further increased the first higher-order ${\rm TE}$ mode starts to break through with the single-mode cutoff width being $w_{\rm co} \simeq 600\,{\rm nm}$ at the design wavelength.

\begin{figure}[htbp!]
\centering
\includegraphics[width=0.8\columnwidth]{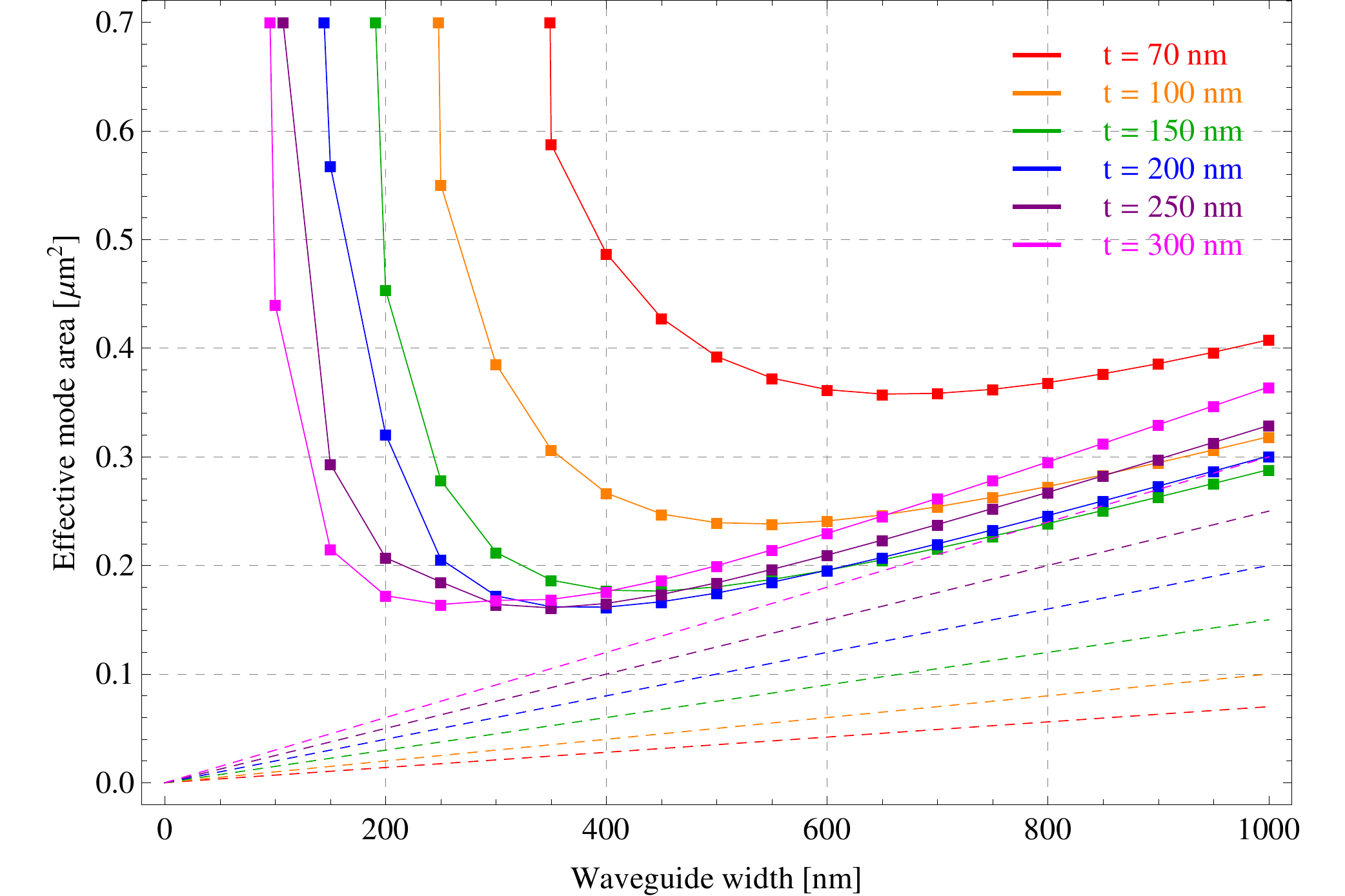}
\caption{Simulated ($\lambda=850\,{\rm nm}$) effective mode area as function of waveguide width for different waveguide thicknesses. Dashed lines show the nonlinear core area $a_{\rm NL}$ which, according to the definition in (\ref{eqn: effective mode area}), is the minimal effective mode area for a given cross section.}
\label{Fig: Effective mode area}
\end{figure}

An equally important property related to the waveguide cross section is the effective mode area $A_{\rm eff}$ introduced in section~\ref{sec: Kerr nonlinearity}. Since the Hamiltonian coupling constant (Eq.~\ref{eq: coupling constant}) is inversely proportional to $A_{\rm eff}$, minimization of this parameter will have a direct impact on the device efficiency. Different definitions of $A_{\rm eff}$ exist in the litterature, but here we adopt the form introduced by Rukhlenko et al.~\cite{Rukhlenko2012} given by
\begin{equation}
A_{\rm eff}=\left. a_{\rm NL}\int\!\!\!\!\int_{-\infty}^{\infty} S_{\rm z} d x d y \middle/ \int\!\!\!\!\int_{\rm NL} S_{\rm z}  d x d y \right. ,
\label{eqn: effective mode area}
\end{equation} 
where $a_{\rm NL}$ is the cross sectional area of the nonlinear waveguide core and $S_{\rm z}$ is the time-averaged $z$ component of the Poynting vector, with the $z$-axis defined to be along the propagation direction of the bus waveguide. The effective mode area is thus simply defined as the core area scaled by the ratio of total mode power to power transmitted through the core. Simulated effective ${\rm TE_0}$ mode areas as function of waveguide width are shown in Fig.~\ref{Fig: Effective mode area} for a range of different thicknesses. 

Based on the simulations in Figs.~\ref{Fig: Effective mode index}-\ref{Fig: Effective mode area} and the imposed fabricational constraints we conclude that a cross sectional waveguide geometry with $t=250\,{\rm nm}$ and $w=500\,{\rm nm}$ would be a favourable choice meeting the requirements of being single mode, polarization maintaining, and having a small effective mode area. Further simulations will henceforth be based on these parameter values. 

\subsection{RTR coupling efficiency}
We now turn to a discussion of the RTR to bus waveguide coupling efficiency and the importance of tailoring this parameter for the application of squeezed light generation. In many applications of micro resonators for nonlinear optics it is merely required that the circulating power be as large as possible, in order to maximize the nonlinear effects. In that case critical coupling is the targeted design. However, as briefly mentioned in section \ref{sec: the system}, it is as crucial for the application at hand that the cavity field predominantly decays through coupling to the bus waveguide and not via intracavity loss channels. The importance lies in the fact that any loss from the quantum correlated field is necessarily accompanied by admixing of vacuum fluctuations which will rapidly outbalance the induced quantum noise reduction and increase the noise floor back to the shot noise level. The figure of merit is the escape efficiency $\eta_{\rm esc}=\gamma_{\rm c}/\gamma$ which should be as large as possible to ensure overcoupling and efficient routing of the cavity field, but at the same time a significant field enhancement factor of the cavity should be maintained.

As mentioned, the intrinsic cavity loss rate $\gamma_0$ is primarily due to scattering losses, and for buried channel silicon nitride waveguides a conservative estimate for the propagation loss parameter is $\alpha \simeq -2\,{\rm dB/cm}$. A second contribution to the overall loss are bending losses due to sub-total internal reflection in waveguide bendings with small radii of curvature. To estimate the impact of this on the overall cavity loss, we have performed simulations of the bending loss suffered by the fundamental TE mode for different radii of curvature. The simulations are based on a combination of a 2D vectorial mode solver and coupled mode theory. Our findings are summarized in Table~\ref{tb: bending loss} and we observe that the effect of this loss channel can readily be reduced one order of magnitude below the linear propagation loss by increasing the resonator bending radius, while still maintaining a footprint of the structure  well below $1\,{\rm mm^2}$. Finally, a third source of optical loss in RTRs are the transitions between straight and bending segments of the resonator. At these locations the optical field experiences abrupt changes in effective refractive index, acting as scattering centers within the waveguide. We will not pursue this loss mechanism further but just mention that the effect can be mitigated by bridging the transitions with adiabatically bending segments at the expense of an increased total resonator round trip length~\cite{Chen2012,Bogaerts2011}.
\begin{table}
\centering
\caption{\label{tb: bending loss}Simulated ($\lambda=850\,{\rm nm}$, $t=250\,{\rm nm}$, and $w=500\,{\rm nm}$) bending loss for the ${\rm TE}_0$ mode in resonators with different radii of curvature, $R$.}
\begin{tabular}{lll}
\hline
R & \multicolumn{2}{c}{Simulated bending loss} \\

[$\mu m$] & [dB/360$^{\circ}$] & [dB/cm] \\
\hline
25 & -1.62$\cdot 10^{-3}$ & -0.103 \\
50 & -3.07$\cdot 10^{-3}$ & -0.0977 \\
75 & -4.35$\cdot 10^{-3}$ & -0.0923 \\
100 & -5.59$\cdot 10^{-3}$ & -0.0890 \\
\hline
\end{tabular}
\end{table}

\begin{figure}[htbp!]
\centering
\includegraphics[width=0.95\columnwidth]{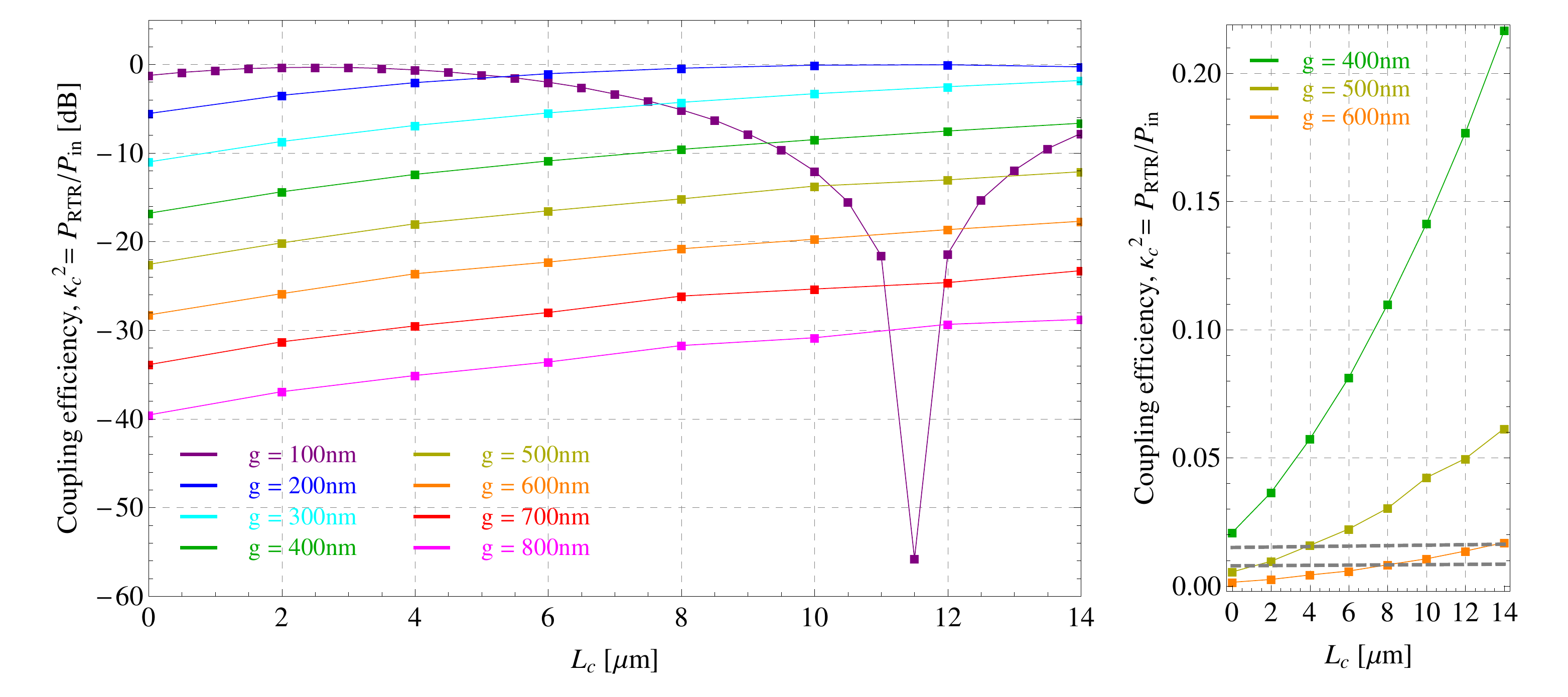}
\caption{Simulated ($\lambda=850\,{\rm nm}$, t = 250\,{\rm nm}, w = 500\,{\rm nm}) efficiency of ${\rm TE}_0$ coupling between RTR and bus waveguide as function of coupling length $L_{\rm c}$. (left) Simulation runs with gap sizes in the range $100 - 800\,{\rm nm}$. (right) Selected simulation runs plotted on linear scale. Gray dashed lines indicate the total intra cavity loss for an RTR with $R=50\mu m$ in case of -1 dB/cm (lower) and -2 dB/cm (upper) propagation loss.
}
\label{Fig: RTR coupling efficiency}
\end{figure}

Based on the above estimates the total round-trip loss is expected to be of the order of a few percent. To achieve strong overcoupling this has to be over-compensated for by the incoupling efficiency, which in the case of an RTR geometry is controlled both through the gap size $g$ and the length $L_{\rm c}$ of the coupling region (see Fig.~\ref{Fig: RTR_topview}). Figure~\ref{Fig: RTR coupling efficiency} shows the simulated dependence of the coupling efficiency on $g$ and $L_{\rm c}$. For large gaps the mode overlap is too small to achieve efficient in and out coupling to the resonator whereas for very small gaps the energy oscillates back and forth between bus waveguide and RTR as function of the coupling length. Furthermore, we observe that for $L_{\rm c}=0\,\mu m$ (ring resonator) the coupling efficiency drops off exponentially with gap size. Relevant simulation runs complying with the constraint on lateral feature sizes are plotted in the righthand panel. 
It appears that for $g=500\,{\rm nm}$ the coupling efficiency can be conveniently tuned across and above the full range of expected intracavity loss values (gray dashed lines), rendering this gap size a favourable choice for fabrication of overcoupled resonators. Reducing the gap by another 100\,nm leads to a strong perturbation of the intrinsic RTR properties and the associated increase in $\eta_{\rm esc}$ comes at a price of a much reduced finesse. Denoting the total intra-cavity power loss by $\mathcal{L}$ we can write approximate expressions for the RTR finesse and escape efficiency,
\begin{equation}
\mathcal{F} \approx \frac{2\pi}{\kappa_{\rm c}^2 + \mathcal{L}}, \qquad \qquad \qquad \eta_{\rm esc} \approx \frac{\kappa_{\rm c}^2}{\kappa_{\rm c}^2 +\mathcal{L}},
\label{eqn: RTR params}
\end{equation}
valid for $\kappa_{\rm c}^2,\, \mathcal{L} \ll 1$. Table~\ref{tb: RTR coupling} summarizes the evaluation of (\ref{eqn: RTR params}) using simulation data from Fig.~\ref{Fig: RTR coupling efficiency}. We observe that for $g=500\,{\rm nm}$ it is indeed possible to achieve escape efficiencies above 80\% and maintain $\mathcal{F}>100$, provided that waveguide propagation losses can be reduced to about -1 dB/cm.

\begin{table}
\centering
\caption{\label{tb: RTR coupling} Evaluation of RTR ($R=50\,{\rm \mu m}$) escape efficiency and finesse in case of -1~dB/cm (left) and -2~dB/cm (right) propagation loss. Italicized values indicate that the approximations used in (\ref{eqn: RTR params}) are only partially justified.}
\begin{tabular}{llrlr}
\hline
$L_{\rm c}$ & \multicolumn{2}{c}{g = 400 nm} &  \multicolumn{2}{c}{g = 500 nm} \\

[${\rm \mu m}$] & $\eta_{\rm esc}$ & $\mathcal{F}$ & $\eta_{\rm esc}$ & $\mathcal{F}$ \\
\hline
4 & 0.88 & 96 & 0.73 & 207\\
6 & 0.91 & 70 & 0.79 & 163\\
10 & \textit{0.94} & \textit{42} & 0.84 & 124\\
14 & \textit{0.96} & \textit{28} & 0.88 & 90\\
\hline
\end{tabular}
\qquad
\begin{tabular}{llrlr}
\hline
$L_{\rm c}$ & \multicolumn{2}{c}{g = 400 nm} &  \multicolumn{2}{c}{g = 500 nm} \\

[${\rm \mu m}$] & $\eta_{\rm esc}$ & $\mathcal{F}$ & $\eta_{\rm esc}$ & $\mathcal{F}$ \\
\hline
4 & 0.79 & 86 & 0.59 & 167 \\
6 & 0.84 & 65 & 0.66 & 137  \\
10 & \textit{0.90} & \textit{40} & 0.73 & 108 \\
14 & \textit{0.93} & \textit{27} & 0.79 & 81  \\
\hline
\end{tabular}
\label{tb: RTR coupling}
\end{table}

\subsection{Robust chip coupling}
The notoriously large coupling loss associated with integrated waveguides imposes further severe limitations on the collection efficiency of the device. In this section we consider a waveguide design based on the TriPleX\texttrademark \,waveguide technology developed by LioniX and we investigate numerically to what extend the design can improve on chip interfacing. Implementation of non-standard designs will inevitably increase fabrication costs and difficulties considerably but it is paramount that the coupling loss issue is addressed thoroughly in order for integrated quantum optical devices to be of any practical relevance.

\begin{figure}[htbp!]
\centering
\includegraphics[width=0.9\columnwidth]{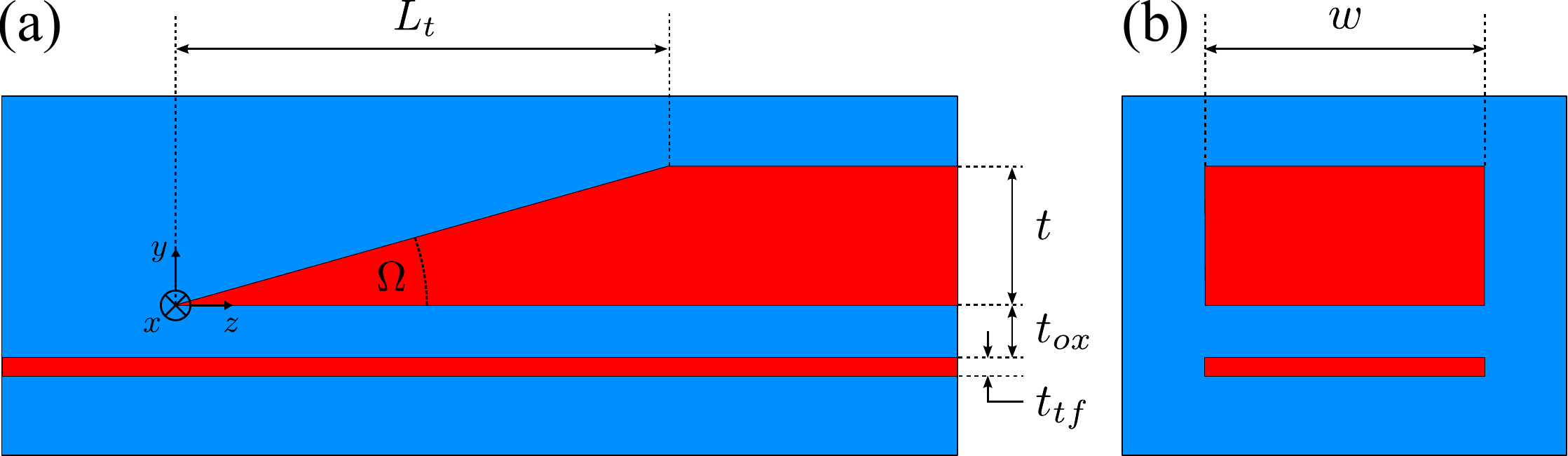}
\caption{(a) Longitudinal cross section of the proposed silicon nitride double layer stack and inverse taper for robust chip coupling. Silicon nitride is indicated in red and the surrounding oxide cladding in blue. The performance of the design depends on optimization of thin film thickness $t_{tf}$, intermediate oxide layer thickness $t_{ox}$, and the tapering length $L_t$. (b)~Transverse cross section of the double layer stack. The lower thin film mirrors the lithographically defined pattern of the upper main waveguide, but with the essential exception that it extends all the way to the chip facets, acting as a weakly guiding structure in connection with both in- and out-coupling.}
\label{Fig: TaperCrossSection}
\end{figure}

The design in question is sketched in Fig.~\ref{Fig: TaperCrossSection} and consists of an asymmetric silicon nitride double layer stack with inverse vertical tapering of the upper layer. The lower film is defined by the same lithographic process as the upper layer, but unlike the latter it extends to the chip facet and acts as a weakly guiding structure for the incident light field serving two purposes: (i) it reduces interface Fresnel losses to $\sim\!\!\,4\%$ ($\eta_{\rm Fresnel} = 96\%$), which can in principle be further reduced by anti-reflection coating, and (ii) the mode field diameter of the guided thin film mode can be easily tailored to match that of an incoming mode delivered from a tapered lensed fibre by controlling the film thickness. Light is then transferred to the main high-confinement waveguide by means of the inverse vertical taper. The efficiency of this process relies on optimization of the intermediate oxide layer thickness and the tapering angle $\Omega$ being sufficiently shallow for the mode conversion to take place adiabatically.

\begin{figure}[htbp!]
\centering
\includegraphics[width=\columnwidth]{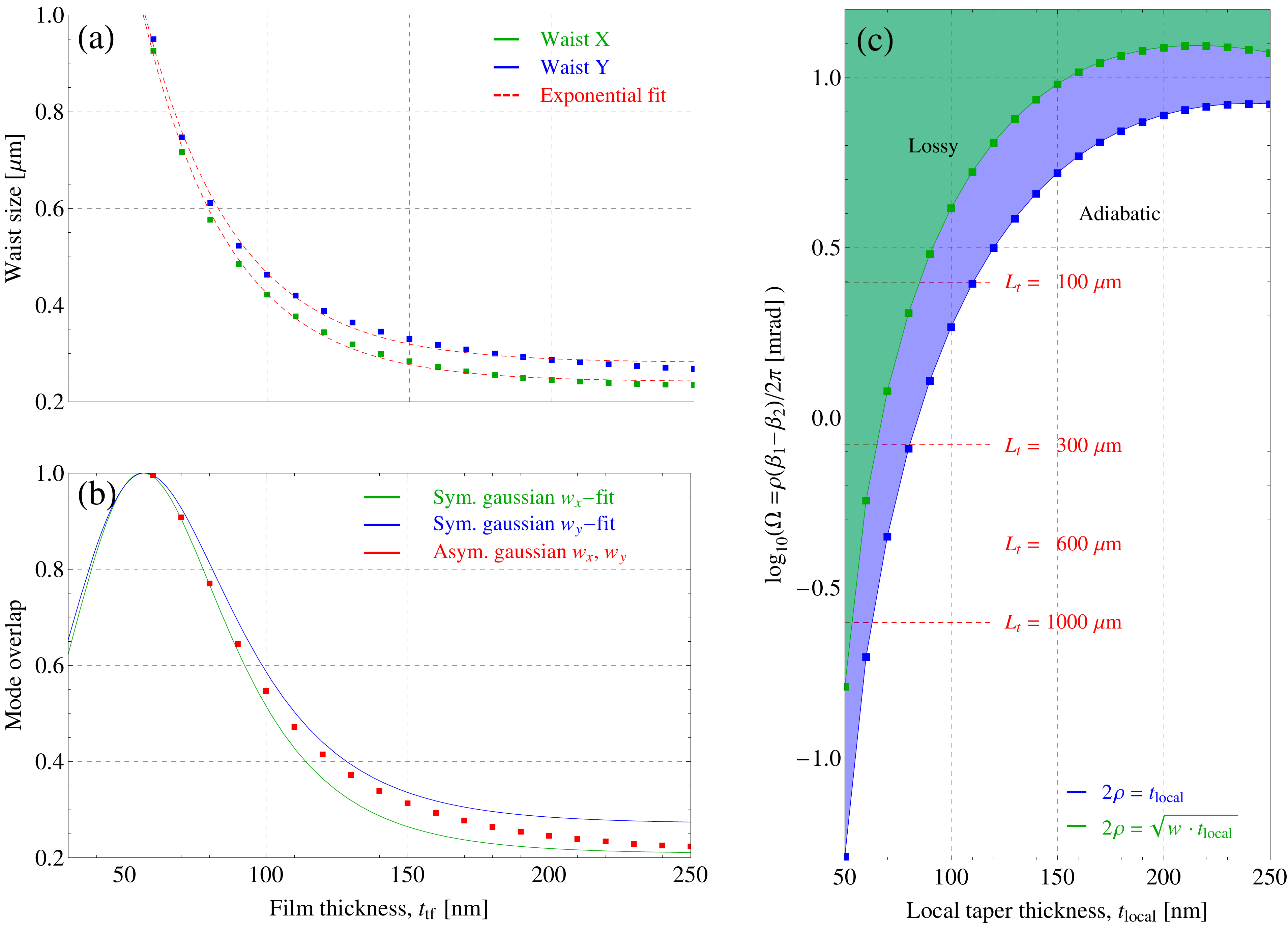} 
\caption{Simulated ($\lambda=850\,{\rm nm}$, w = 500\,{\rm nm}) mode waist sizes along major (X) and minor (Y) axes of the lower stripe waveguide as function of film thickness (a), and corresponding overlap efficiency with an incoming symmetric Gaussian mode with a waist of $1\,\mu m$ (b). (c) Simulated delineation curves for approximate taper adiabaticity. Dashed horizontal lines correspond to linear constant-slope tapers of varying taper length.}
\label{Fig: StripeMode}
\end{figure}

Based on mode analysis simulations of the fundamental TE mode of the lower stripe waveguide as function of film thickness, we plot in Fig.~\ref{Fig: StripeMode}a the resulting mode waist size along with exponential fits. The resulting overlap between the asymmetric waveguide mode and a symmetric incoming Gaussian mode with a waist of $1\,\mu m$, consistent with lensed tapered fibre coupling, is evaluated and plotted in Fig.~\ref{Fig: StripeMode}b (squares).  For comparison, we also evaluate the closed form expression for the overlap integral of two symmetric Gaussian beams (waists $w_1, w_2$)~\cite{Pollock2010} \begin{equation}
\eta = \frac{4 w_1^2 w_2^2}{(w_1^2 + w_2^2)^2},
\end{equation}
with $w_1 = 1\,\mu m$ and the fitted exponential functions for $w_x$ and $w_y$ substituted for $w_2$,  respectively. Our calculations suggest that an overlap efficiency of $\eta_{\rm overlap} \geq 98\%$ is achievable for a film thickness of 60 nm. For comparison, butt coupling a standard single mode fibre, with mode field diameter of $5\,\mu m$, to the untapered waveguide would result in an unfeasible mode overlap of only 2\%. 

As can be seen from Fig.~\ref{Fig: StripeMode}b the mode overlap is a rather peaked function of film thickness, because of the exponentially increasing mode waist, meaning that the film thickness should be well controlled during fabrication. Using only a single silicon nitride layer the etching process defining the inverse taper must be terminated exactly at the targeted remaining film thickness, requiring a homogeneous and carefully calibrated etch rate. Furthermore, the etching process would result in considerable surface roughness of the thin film top surface causing increased scattering losses. The proposed double layer stack design conveniently circumvents these fabricational issues: the thin film thickness is well-controlled through the LPCVD deposition process and the constraint on etch depth control can be relaxed because of the protective intermediate oxide layer.  

We now turn to the transfer of power from the thin lower stripe to the main waveguide. In order to put contraints on the waveguide taper design we adopt the theory developed by Love et al.~\cite{Love1991,Ladouceur1996} for delineation of adiabaticity criteria for tapered optical fibres. Following this treatment, approximately adiabatic propagation of the fundamental waveguide mode along the taper is ensured if the local taper length-scale $z_{\rm t}$ is much larger than the coupling length $z_{\rm b} =2\pi/(\beta_1 -\beta_2)$ between the fundamental and second local modes everywhere along the taper. Here $\beta_i = n_{\rm eff}(\omega_{\rm i}) k_{\rm 0,i}$ are the propagation constants for the involved modes, and for tapering angles $\Omega \ll 1$ the local taper length-scale is $z_{\rm t} \approx \rho/\Omega$, where $\rho$ is a local characteristic transverse length scale. For the linear constant-slope tapers cosidered here, $z_{\rm t}$ is always equal to the position $z$ along the taper. The condition $z_{\rm t} = z_{\rm b}$ leads to the approximate delineation criterion
\begin{equation}
\Omega = \frac{\rho(\beta_1 -\beta_2)}{2\pi}.
\end{equation}      
For angles much larger than $\Omega$ a significant amount of power is coupled to the second local mode and the taper becomes lossy. Figure~\ref{Fig: StripeMode}c shows delineation curves based on numerical mode analyses of the ${\rm TE}_0$ and ${\rm TM}_0$ modes along the taper. The adiabaticity criterion has been evaluated using both the local taper thickness $t_{\rm local}$ and the geometrical mean of the local cross sectional dimensions as characteristic length scale. Horizontal dashed lines indicate tapering angles corresponding to constant-slope tapers of length in the range $100 - 1000\,\mu m$. It is obvious from the figure that any constant-slope taper will eventually enter the non-adiabatic region. However, by making the taper sufficiently long the associated loss can be reduced to a low level. In the lossy region approximately 10\% of the power is coupled to the second mode per half coupling length, and as $z_{\rm b}$ increases dramatically as the taper thins down we estimate that the coupling loss can be reduced to $\leq \!10\%$ for realistic taper lengths of $500-600\,\mu m$. 
To consolidate this estimate, we have simulated the power transfer between lower thin film and main waveguide using 2D BPM simulations with the target parameter values derived in this section and using different intermediate oxide layer thicknesses.  Without going further into details we mention that the simulation results suggest that a total transfer efficiency of the taper exceeding $\eta_{\rm taper}= 90\%$ is indeed realizable with an oxide thickness of $t_{\rm ox} \approx 100\,{\rm nm}$.

In summary, the above analysis yields a promising estimate for the total per facet coupling efficiency for the asymmetric double layer stack design of
\begin{equation}
\eta_{\rm c} = \eta_{\rm Fresnel} \cdot \eta_{\rm overlap} \cdot \eta_{\rm taper} \geq 0.96 \cdot 0.98 \cdot 0.90 \approx 84\%,
\end{equation}
corresponding to -0.75 dB. Experimental confirmation of this value would be an important step towards feasible integrated high-index contrast structures for quadrature squeezed light generation.

\section{Source characterization}
\label{sec: results}
The figure of merit characterizing the feasibility of the proposed integrated device is ultimately the maximum achievable quantum noise reduction. But other characteristica such as output state purity and bandwidth are also important for the practical applicability of the source and will as well be discussed in the following. 

In an earlier study of squeezing from cavity enhanced third-order nonlinearities by S. Reynaud et al.~\cite{Reynaud1989} it was pointed out that perfect amplitude squeezing is, in principle, obtainable close to the turning points of the bistability curve (points on the black dashed contour in Fig.~\ref{Fig: PumpEigenvalue}). However, in the present study we will restrict our evaluation to the case where the pump field is adaptively kept on resonance with the hot cavity as the incident power in the bus waveguide is increased (dashed red line in Fig.~\ref{Fig: PumpEigenvalue}). In this way the intra-cavity power is maximized and the steady state behaviour of the system is reduced to that of an empty cavity, deliberately eliminating optical bistability. 

The simulation results of section~\ref{sec: numsim} provides a set of favorable target parameter values for the RTR design and the necessary inputs for estimating the resulting nonlinear interaction strength. Leaving the RTR power coupling efficiency as the only variable design parameter, we proceed to evaluate the theoretical quantum noise spectrum (\ref{eqn: pump spectrum}-\ref{eqn: effective squeezing spectrum}) as function of experimentally accessible parameters external to the integrated device, yielding realistic values for key characteristics of the device. 
We assume a $250 \times 500$\,nm waveguide cross section for which Fig.~\ref{Fig: Effective mode area} yields an effective mode area of $A_{\rm eff} = 0.18\, \mu m^2$ corresponding to a nonlinear parameter of $\gamma_{\rm nl} = 10.2\,{\rm W^{-1}m^{-1}}$. In the following, we further assume linear propagation losses to be -1\,dB/cm and an RTR coupling gap of 500\,nm. In order to accommodate the simultaneous requirements of high escape efficiency and appreciable intra-cavity field enhancement, we restrict the coupling efficiency to values in the range 2-8\%. Since the associated modification to the nonlinear interaction strength $\xi$, through the $L_{\rm c}$ dependence of the RTR round-trip length, is only on the few-percent level, we will neglect this and simply assume a constant value given by the corresponding value for $L_{\rm c} = 12\,\mu m$ ($\kappa_{\rm c}^2 = 5\%$), namely $\xi=117\,{\rm Hz}$. 

In order to estimate the measurable degree of squeezing the finite detection efficiency (\ref{eqn: effective squeezing spectrum}) of the setup must also be accounted for. In the previous section, a simulation based estimate for the per facet fibre-chip coupling efficiency was found to be $\eta_{\rm c} \approx 84\%$, and experimentally realistic values for the homodyne mode overlap and detector quantum efficiency are $\eta_{\rm mm} = \eta_{\rm qe} = 98\%$. In total this amounts to a detection efficiency of $\eta = 80.6\%$ which will be assumed in the following unless otherwise stated.  

\subsection{Quantum noise tomography}
\begin{figure}[htbp!]
\center
\includegraphics[width=1\columnwidth]{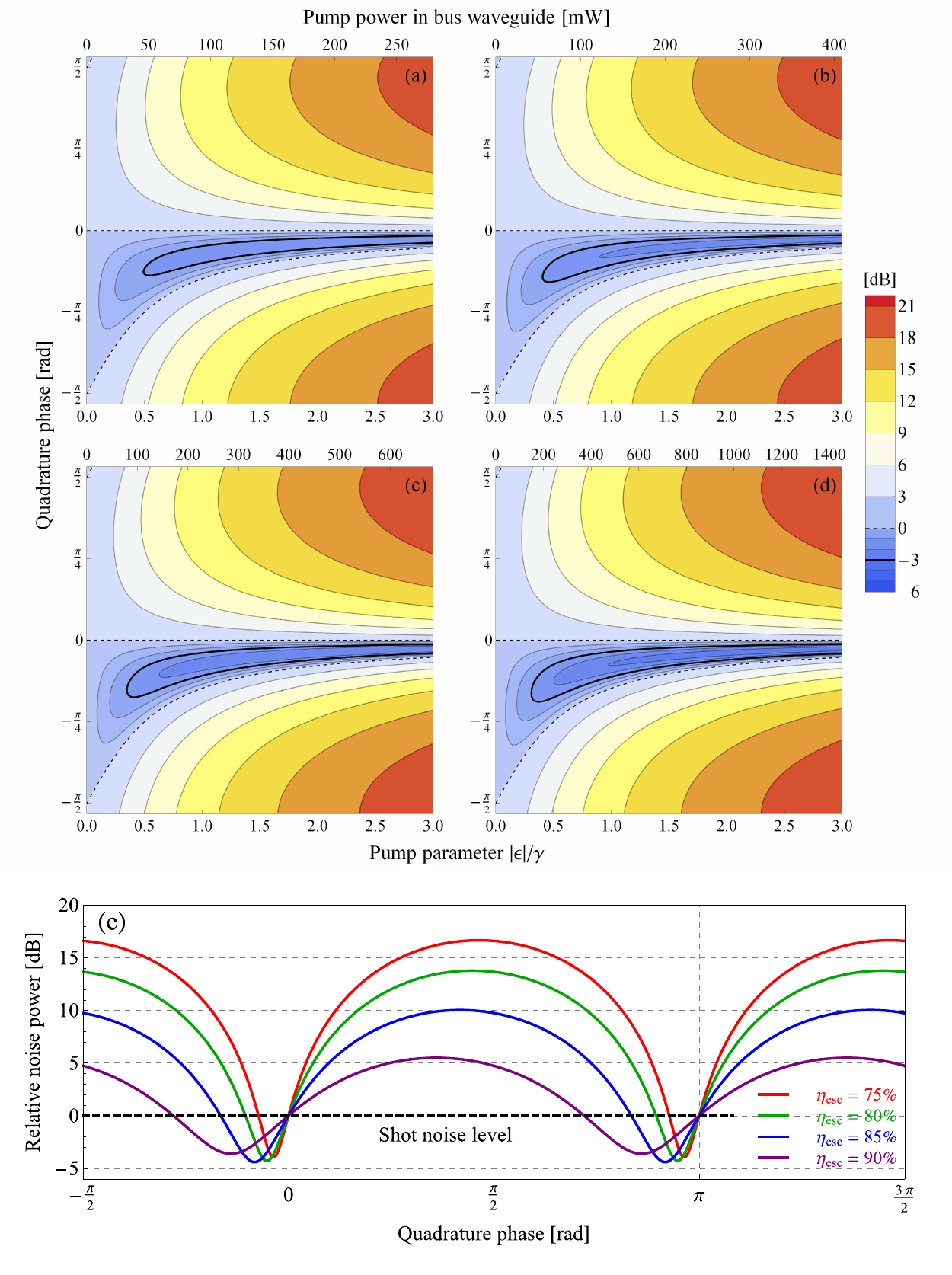}
\caption{($\lambda = 850\,{\rm nm}$, $\alpha = -1\,{\rm dB/cm}$, $\eta$ = 80.6\%, $\Omega/2\pi = 30\,{\rm MHz}$) Relative noise power in the output field of a resonantly pumped RTR ($\Delta_{\rm p} = |\epsilon|$) as function of pump parameter $|\epsilon|$. Panels (a)-(d) correspond to RTR escape efficiencies of 75\%, 80\%, 85\%, and 90\% respectively. Shot noise level is indicated by the dashed contour and the solid black contour corresponds to $-3\,{\rm dB}$ quantum noise reduction. (e) Cross sectional views of panels (a)-(d) corresponding to an input pump power of 200 mW.}
\label{Fig: ResonantPumpSqueezing}
\end{figure}

Using (\ref{eqn: pump spectrum}-\ref{eqn: effective squeezing spectrum}) we calculate the quantum noise characteristics of the output field, accounting for imperfect collection and detection efficiencies. The corresponding relative noise power tomography is plotted in Fig.~\ref{Fig: ResonantPumpSqueezing} as function of pump power for increasing escape efficiencies, assuming a particular sideband frequency of $\Omega/2\pi = 30\,{\rm MHz}$ well above any technical noise in a standard experimental setting. Most significantly, a quantum noise reduction of about -4.5\,dB is observable in the output field, confirming the feasibility of the designed device as a source of bright squeezing. 
Characteristic to the Kerr effect, both squeezing efficiency and phase depend on pump power and in conformity with the input-output relations (\ref{eq: Kerr input-output}) the amplitude quadrature noise is unaffected by the interaction.

As previously discussed, it is generally of interest to maximize the escape efficiency of a resonant source to preserve squeezing in the output field. However, for the system in question, design optimization is not as straight forward due to the self-driven squeezing interaction and power dependence of the Kerr nonlinearity: increasing $\eta_{\rm esc}$ inevitably reduces the cavity finesse and thereby the intra-cavity power driving the squeezing process. We observe from Fig.~\ref{Fig: ResonantPumpSqueezing} that for a given $|\epsilon|$ (bottom axes) larger squeezing is indeed achievable as the escape efficiency is increased. But the required pump power in the bus waveguide (top axes) increases accordingly, rapidly approaching experimentally intractable levels. For $\eta_{\rm esc} \geq 90\%$ we find that reaching -3 dB of squeezing already requires about 200 mW input pump power. 

\subsection{State purity}
\begin{figure}[htbp!]
\centering
\includegraphics[width=1\columnwidth]{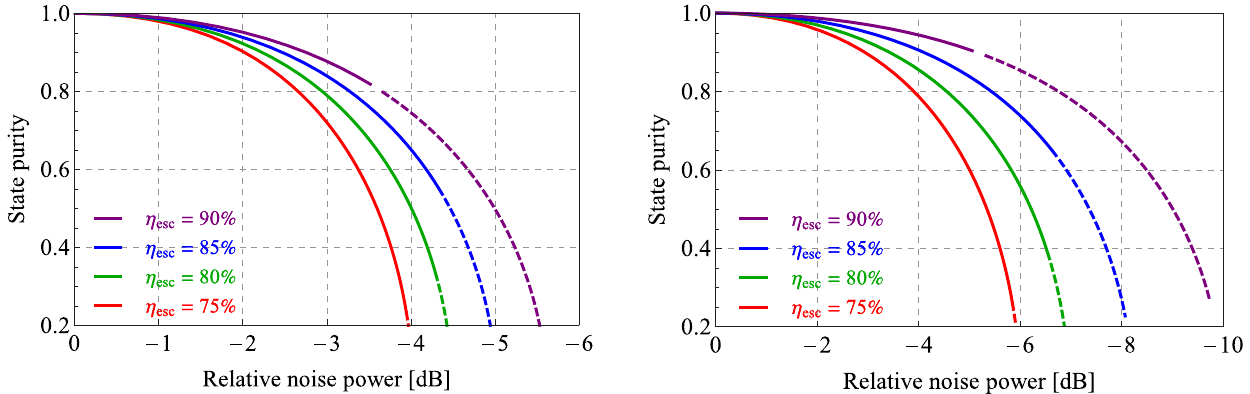}
\caption{State purity as function of noise reduction for different RTR escape efficiencies. Dashed segments indicate that generation of the corresponding noise reduction requires a pump power in the bus waveguide exceeding 200\,mW. The two panels correspond to detection efficiencies of $\eta$ = 80.6\% (left) and $\eta = 100\%$ (right).}
\label{Fig: ResonantPumpSqueezingPurity}
\end{figure}

\begin{figure}[htbp!]
\center
\includegraphics[width=0.48\columnwidth]{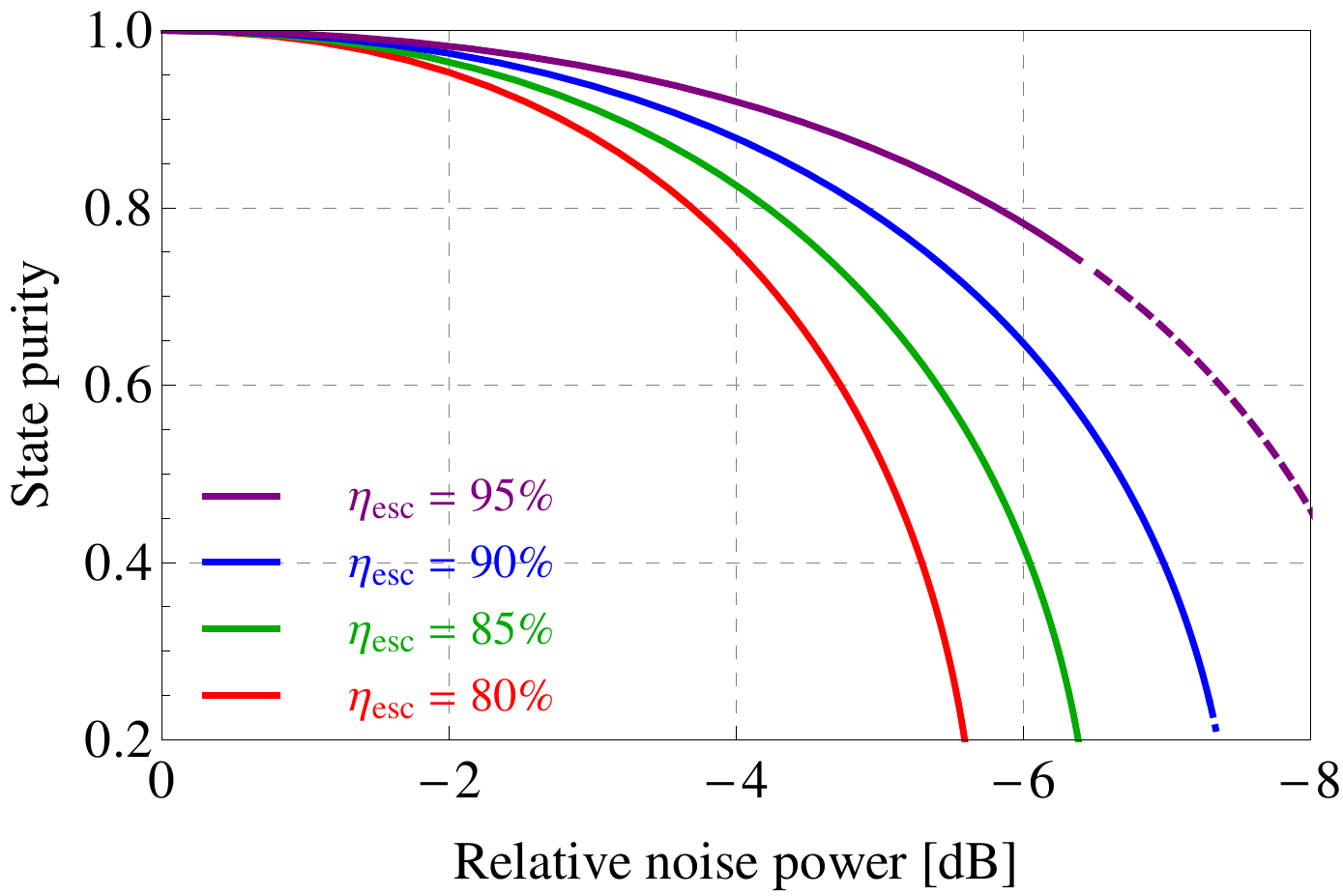}
\caption{State purity as function of noise reduction with device loss properties of $\alpha$ = -0.2 dB/cm and $\eta_{\rm c} = 95\%$. Dashed segments indicate that generation of the corresponding noise reduction requires a pump power in the bus waveguide exceeding 50\,mW.}
\label{Fig: ResonantPumpSqueezingPurity_lowloss}
\end{figure}

From the tomography in Fig.~\ref{Fig: ResonantPumpSqueezing} we note that the anti-squeezing rises steeply with pump power compared to the squeezing level. To further characterize this property, we evaluate the state purity, which for Gaussian states is defined as
$\mu = 1/\sqrt{{\rm Det}[\boldsymbol\sigma]}$ \cite{Paris2003}, $\boldsymbol\sigma$ being the covariance matrix for the state. Subject to optical loss, the purity of a squeezed state will drop quickly due to thermalization with the vacuum, rendering it practically impossible to produce highly squeezed minimum uncertainty states saturating the Heisenberg inequality, ${\rm Var}(\delta\!X_1) \cdot {\rm Var}(\delta\!X_2) = 1$. Due to the comparatively large intra-cavity and coupling loss associated with the RTR source, considerable excess anti-squeezing and correspondingly low state purity is anticipated even for modest squeezing levels. Using the simulated data in Fig.~\ref{Fig: ResonantPumpSqueezing} we evaluate the state purity as function of squeezing, plotted in Fig.~\ref{Fig: ResonantPumpSqueezingPurity}, both for the estimated experimental detection efficiency and in the idealized case of unity efficiency.
As expected, the purity increases with $\eta_{\rm esc}$ but at the expense of large pump powers. To circumvent this, a lower intra-cavity loss rate is required. Furthermore, Fig.~\ref{Fig: ResonantPumpSqueezingPurity} (right) shows that despite the efforts to minimize fibre-chip coupling losses the device collection efficiency remains a limiting issue. It is clear that the inherent loss channels of the source pose a severe limitation to the performance which can only be overcome by improved fabricational quality of the device. For comparison we plot in Fig.~\ref{Fig: ResonantPumpSqueezingPurity_lowloss} the performance of the device with waveguide propagation loss and  collection efficiency set to -0.2\,dB/cm and 95\%, respectively. This allows for an increase in escape efficiency and a dramatic reduction of the required pump power, enabling -6\,dB of squeezing with a purity of $\mu \approx 0.8$ using less than 50\,mW pump power.

\subsection{Squeezing bandwidth}
Finally, we investigate the noise spectrum of the RTR output field. Commonly, cw parametric down-conversion sources of quadrature squeezed light are limited to squeezing bandwidths of a few tens of MHz, set by the OPO cavity bandwidth. One exception being monolithic cavity systems (\cite{Mehmet2010b,Yonezawa2010}) where the internal loss is sufficiently low to reach bandwidths up to 170\,MHz. Weak squeezing with a bandwidth of more than  2\,GHz at 1550 nm has also been demonstrated in such a system solely resonant for the pump field~\cite{Ast2012}, and in a later work about 2 dB of squeezing was observed with a bandwidth exceeding 1\,GHz~\cite{Ast2013}. 

Squeezed light is a widely used resource in quantum communication applications, e.g. for establishing the essential two-partite entanglement for quantum key distribution \cite{Cerf2001,Madsen2012}. In this particular case the key rate is directly linked to the squeezing bandwidth and demonstration of efficient GHz-bandwidth sources is thus an important prerequisite for high-speed protocols. Large-bandwidth squeezed states are also essential for extending the applicability of cv quantum sensing techniques. As an example, state of the art optomechanical photonic crystal devices have mechanical resonance frequencies in the GHz range~\cite{Safavi-Naeini2012}, rendering conventional squeezed light sources insufficient for enhancing the transduction sensitivity. The integrated source design discussed in this paper has potential to fulfill those requirements. As shown in the preceding sections, efficient squeezing is indeed possible and due to the intrinsic phase matching of the SPM process the spectral extend of the squeezing is limited by the cavity bandwidth alone. 

\begin{figure}[htbp!]
\centering
\includegraphics[width=0.98\columnwidth]{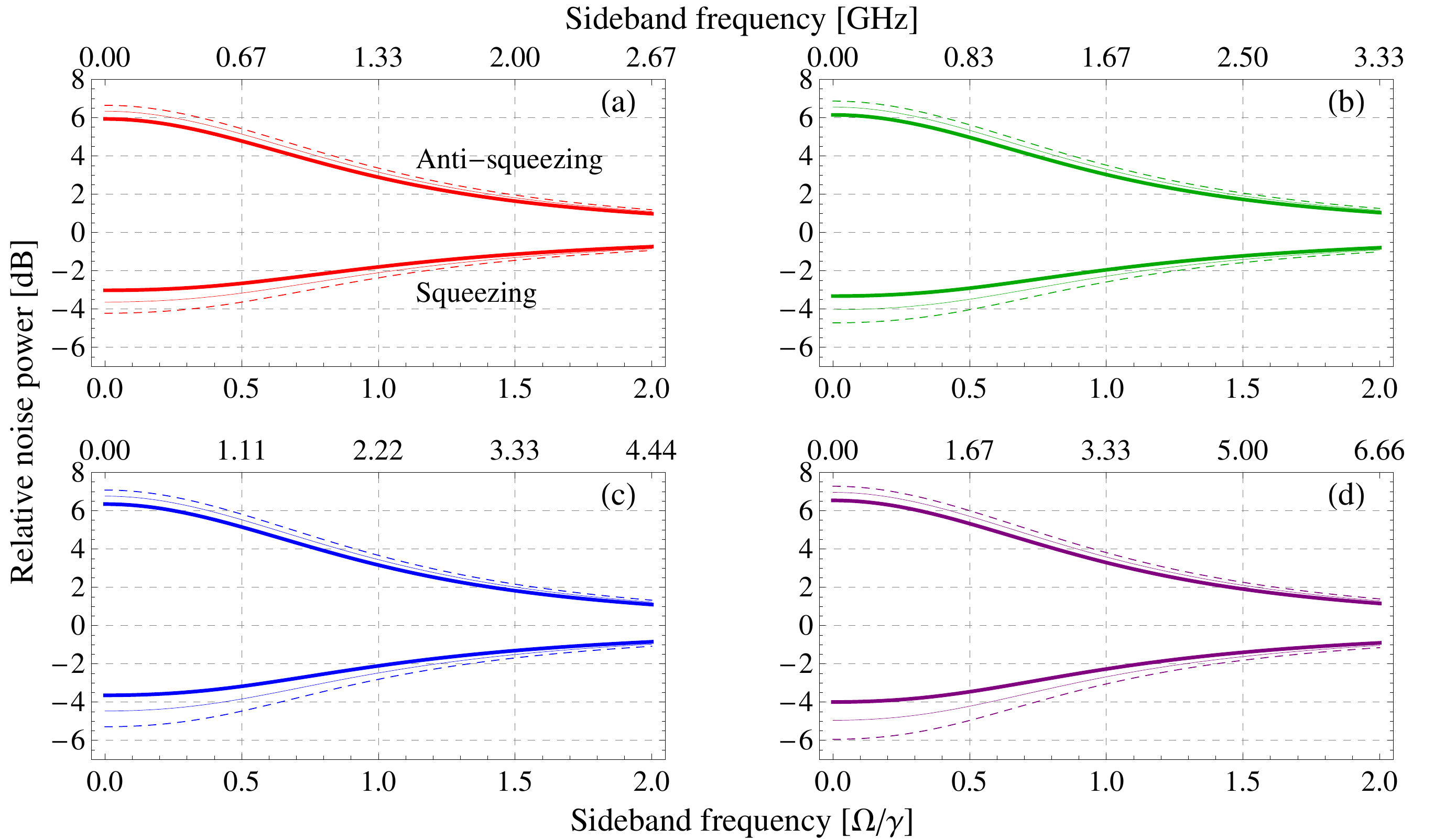} 
\caption{($\lambda = 850\,{\rm nm}$, $\alpha = -1\,{\rm dB/cm}$, $|\epsilon|/\gamma = 0.5$) Spectrum of the squeezed and anti-squeezed quadratures assuming 80.6\% (thick), 91.2\% (thin), and unity (dashed) detection efficiency.  Panels (a)-(d) correspond to RTR escape efficiencies of 75\%, 80\%, 85\%, and 90\% respectively.}
\label{Fig: ResonantPumpSqueezingSpectra}
\end{figure}
Figure~\ref{Fig: ResonantPumpSqueezingSpectra} shows the spectra of the squeezed and anti-squeezed quadratures corresponding to $|\epsilon|/\gamma = 0.5$ and for the same escape efficiencies as in Figs.~\ref{Fig: ResonantPumpSqueezing} and~\ref{Fig: ResonantPumpSqueezingPurity}. With a total detection efficiency of 80.6\% ($\eta_{\rm c}$ = 84\%) the output field shows more than -2\,dB quantum noise reduction over a bandwidth larger than 1\,GHz across all four $\eta_{\rm esc}$-values. If the detection efficiency is increased to 91.2\% ($\eta_c$ = 95\%) we expect more than -3 dB squeezing over a bandwidth of 1.5\,GHz for $\eta_{\rm esc} \geq 85\%$.

\section{Conclusion}
While on-chip photonics is already a promising technology gradually evolving into a versatile, modular, and reconfigurable architecture, development of an integrated platform for cv quantum optics has only received little attention. That despite the vast toolbox of Gaussian cv quantum sensing, computation, and communication protocols, already successfully implemented and demonstrated in free-space optical systems, but unsuitable for marrying current technology without a scalable architecture and cost-efficient fabrication line. Squeezed states are one of the most commonly deployed resources for quantum-enhanced sensing applications and entangled-state generation for communication protocols, and as such development and demonstration of an efficient integrated source of quadrature squeezed states is an important step towards a cv state-based quantum-enhanced optical technology. 

We have presented a theoretical feasibility study of a resonant integrated source of bright single-mode squeezed light, exploiting the third-order optical nonlinearity of silicon nitride. Through numerical simulations we have studied the dependence of optical propagation and nonlinear properties on waveguide cross section geometry as well as resonator coupling and chip interfacing. The issue of optical loss in integrated optics and its detrimental influence on optical quantum states has been addressed. To this end we have advocated an asymmetric double layer stack waveguide geometry with inverse vertical tapers as a means for mitigating the inherently large coupling losses associated with interfacing high-confinement waveguides. On basis of the simulations we have proposed a concrete design optimized for generation and collection of squeezed light, and the feasibility of the design was evaluated from the derived squeezing spectrum of the output field. The theoretical predictions are encouraging, suggesting that the source is capable of producing about -4.5\,dB of squeezing with 200 mW pump power, mainly limited by the chip-fibre out-coupling efficiency and under the conservative assumption of -1 dB/cm propagation loss. Furthermore, a quantum noise reduction exceeding -2\,dB is expected over a bandwidth larger than 1\,GHz. If the waveguide propagation loss can be reduced to -0.2\,dB/cm and the chip out-coupling efficiency increased to 95\%, then squeezing efficiencies of about -7\,dB are attainable using only 50\,mW pump power.

%%% START OF APPENDIX %%%
\appendix

\section{Derivation of the Kerr interaction Hamiltonian}
\label{ap: Kerr Hamiltonian}
In this appendix we briefly review the derivation of the quantum mechanical Hamiltonian describing Kerr interaction in a dielectric waveguide. We will limit our analysis to involve only a single spatial mode of the waveguide and linearly polarized monochromatic fields.   

Classically, the field energy in a dielectric medium is given by $\mathcal{H}= \frac{1}{2}\int dV \left\langle \mathcal{E}(\mathbf{r},t) \cdot \mathcal{D}(\mathbf{r},t) \right\rangle $, where the electric displacement is given by $ \mathcal{D}(\mathbf{r},t) = \varepsilon_0  \mathcal{E}(\mathbf{r},t) +  \mathcal{P}(\mathbf{r},t)$ and the brackets denote cycle average. In case of a nonlinear dielectric, the anharmonic response is accounted for by Taylor expansion of the polarization  $\mathcal{P}(\mathbf{r},t)$ in orders of the susceptibility, and for a Kerr medium the lowest non-vanishing contribution is of third order, characterized by the sucecptibility tensor $\chi^{(3)}_{ijkl}$ of rank 4. In general, calculation of the corresponding interaction energy involves the contraction of the susceptibility tensor with four individual fields -- two pump fields ($a_{p1}$, $a_{p2}$) plus upper ($a_+$) and lower ($a_-$) sideband fields.
However, in the case of linearly co-polarized fields the expression simplifies significantly:
\begin{equation}
\mathcal{H}_{\rm I}=3\varepsilon_0 \chi^{(3)}  \frac{1}{2} \int dV \left\langle \mathcal{E}(\mathbf{r},t)^4 \right\rangle,
\label{eqn: Field energy}
\end{equation}
where $ \mathcal{E}(\mathbf{r},t)$ is still assumed to be a superposition of four independent waves. 

Since we are concerned only with a single spatial mode of the waveguide, an appropriate plane-wave expansion of the quantized electric field inside the medium is given by
\begin{equation}
 E(\mathbf{r},t) = i \mathcal{E}_0 \sum_{i}\left( a_{i} e^{i(\mathbf{k}_{i} \cdot \mathbf{r} -\omega_i t)} +  a_i^{\dagger} e^{-i(\mathbf{k}_i \cdot \mathbf{r} -\omega_i t)} \right) \qquad {\rm where} \quad i=\{p_1, p_2, +, - \}
\end{equation}
with normalization constant $\mathcal{E}_0 =\sqrt{\hbar \omega /2 \varepsilon_0 n^2 V}$, where the quantization volume is $V = L \cdot A_{\rm eff}$ given by the product of the effective area of the guidede mode and the waveguide length. We have assumed that the frequencies of the interacting modes are sufficiently close that dispersion effects and the difference in photon energies can be neglected. Replacing $ \mathcal{E}(\mathbf{r},t)$ in (\ref{eqn: Field energy}) by the quantized field operator results in a large number of quadrilinear products of the four involved mode operators, of which all the energy non-conserving terms are readily removed by the cycle average, enforcing the rotating wave approximation. The remaining terms are all permutations of the form $a_i a_j a^{\dagger}_k a^{\dagger}_l$ (and Hermitian conjugates) for which $\omega_i + \omega_j = \omega_k +\omega_l $ and the phase factor associated with these terms is $e^{\pm i(\mathbf{k}_i + \mathbf{k}_j - \mathbf{k}_k -\mathbf{k}_l)\cdot \mathbf{r}} = e^{\pm i\Delta \mathbf{k}\cdot \mathbf{r}} = e^{\pm i\Delta k z}$, where in the last step we have used the assumption of a single mode waveguide. Carrying out the volume integral yields
\begin{equation}
\int d V e^{ i\Delta k z} = \int\int_{A_{\rm eff}} d x d y \int_0^L d z \, e^{ i\Delta k z} = A_{\rm eff} L \Phi (\Delta k L)
\label{eqn: phasematching}
\end{equation}
(or the complex conjugate) where we have defined the phase matching function  $ \Phi (\Delta k L) = e^{i\Delta k L/2} \, {\rm sinc} (\Delta k L/2)$. The effect of wavevector mismatch on energy transfer efficiency of the processes is entirely given by $|\Phi (\Delta k L)|^2 =  {\rm sinc}^2 (\Delta k L/2)$. The factor $e^{i\Delta k L/2}$ only constitutes an immaterial global phase and will be neglected. Using (\ref{eqn: phasematching}) and the expression for $\mathcal{E}_0$ we can express the pre-factor of the quantized Hamiltonian, characterizing the interaction strength $\Xi$:
\begin{eqnarray}
\Xi &=& \frac{1}{2} 3\varepsilon_0 \chi^{(3)}  \left( \sqrt{\frac{\hbar \omega }{2 \varepsilon_0 n^2 V}} \right)^4 A_{\rm eff} L \Phi (\Delta k L)\\
&=& \frac{1}{2} \frac{\hbar^2 \omega c^2 \gamma_{\rm NL}}{n^2 L }\Phi (\Delta k L)\\
&=& \hbar\xi\Phi ( \Delta k L) \quad {\rm where} \quad  \xi =\frac{\hbar \omega c^2 \gamma_{\rm NL}}{2 n^2 L}.
\end{eqnarray}
Here we have used the definition of the nonlinear refractive index $n_2 =3 \chi^{(3)}/4\varepsilon_0 n^2 c$ and the nonlinear parameter
$\gamma_{\rm NL} = n_2 \omega/c A_{\rm eff}$.

\begin{figure}[htbp!]
\centering
\includegraphics[width=0.63\columnwidth]{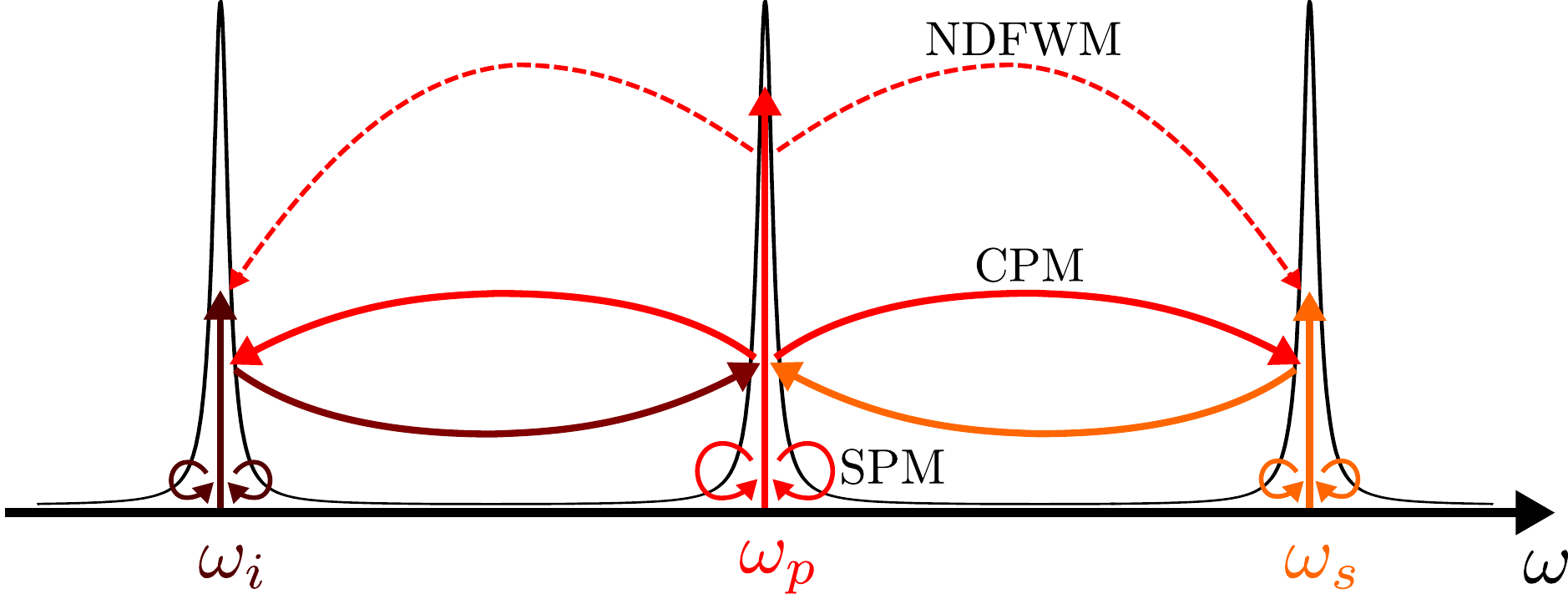}
\caption{Pictorial illustration of the $\chi^{(3)}$-mediated processes SPM, CPM, and NDFWM, in the case where a strongly pumped resonator mode at frequency $\omega_{\rm p}$ interacts with two weaker signal and idler sideband modes at frequency $\omega_{\rm s}$ and $\omega_{\rm i}$, respectively.}
\label{Fig: NLprocesses}
\end{figure}

We now return to the sum of operator products, but we restrict our treatment to the case of degenerate pump fields ($a_{p1} = a_{p2} = a_p$). Furthermore, we choose to write the Hamiltonian in normal ordered form. Following the treatment by  Jack et al.~\cite{Jack1995} the Hamiltonian is then given by
\begin{equation}
H_{\rm I} = 2 \hbar \xi \frac{ :(\sum_i (a_i + a^{\dagger}_i))^4:}{4!} \qquad {\rm where} \quad i=\{p, +, - \}
\label{eqn: App interaction Hamiltonian}
\end{equation}
and we are implicitly assuming that energy non-conserving terms are neglected. The overall factor of 2 accounts for anti-normal ordered terms in the expansion and the 4! in the denominator is for normalization with respect to the total number of permutations. The remaining energy conserving terms and their interpretation are listed in Table~\ref{tb: Hamiltonian terms} below. The interactions described by the three associated processes are sketched in Fig.~\ref{Fig: NLprocesses} in case of resonant interaction between a strong pump field and two weaker signal and idler fields, all resonant on separate longitudinal cavity modes.

\begin{table}[htbp!]
\centering
\caption{ Energy conserving operator products in the expansion of the normal ordered interaction Hamiltonian (\ref{eqn: App interaction Hamiltonian})}
\begin{tabular}{lll}
\hline
Operator terms & Associated physical process  \\
\hline
${a^{\dagger}_{\rm p}}^2 a_{\rm p}^2 \quad {a^{\dagger}_+}^2 a_+^2 \quad {a^{\dagger}_-}^2 a_-^2$ &  Self-phase modulation (SPM) \\
$a_{\rm p}^2 a^{\dagger}_+ a^{\dagger}_-$ & Non-degenerate four-wave mixing (NDFWM) \\
$a^{\dagger}_{\rm p} a_{\rm p} a^{\dagger}_+ a_+ \quad a^{\dagger}_{\rm p} a_{\rm p} a^{\dagger}_- a_- \quad a^{\dagger}_+ a_+ a^{\dagger}_- a_-$  & Cross-phase modulation (CPM) \\
\hline
\end{tabular}
\label{tb: Hamiltonian terms}
\end{table}

From the definition of $\Phi (\Delta k L)$ it is clear that only the FWM proces requires active phase matching whereas both SPM and CPM are intrinsically phase matched. The relative strength of the three processes can be evaluated by using the multinomial distribution  for expanding the Hamiltonian. Calculating the multiplicities for the three processes we get weight factors of 1/2, 1, and 2 for SPM, FWM, amd CPM, respectively. Finally, we can write the total Kerr interaction Hamiltonian (assuming perfect phase mathcing):
\begin{eqnarray}
H_{\rm I} &=& \hbar \xi \Phi (\Delta k L) (a_{\rm p}^2 a_+^{\dagger}  a_-^{\dagger} + {a^{\dagger}_{\rm p}}^2 a_+ a_-)  \\
&& + 2 \hbar \xi (a^{\dagger}_{\rm p} a_{\rm p} a^{\dagger}_+ a_+ + a^{\dagger}_{\rm p} a_{\rm p} a^{\dagger}_- a_- + a^{\dagger}_+ a_+ a^{\dagger}_- a_-) \\
&& + \frac{\hbar \xi}{2}({a^{\dagger}_{\rm p}}^2 a_{\rm p}^2 + {a^{\dagger}_+}^2 a_+^2 + {a^{\dagger}_-}^2 a_-^2).
\end{eqnarray}

\section*{Acknowledgments}
We acknowledge the work of Mathieu Manceau during the early stage of this project. Also, we are grateful for valuable discussions of fabricational matters with Arne Leinse and Erik Schreuder (LioniX BV), and the support of Remco Stoffer (PhoeniX Software BV) on software packages Field Designer and Opto Designer.\\ 
\noindent This work was supported by the Danish Council for Independent Research (Sapere Aude program).

%\bibliography{references}

\end{document}